\documentclass[11pt]{article}

\usepackage{amssymb,amsmath,graphicx}
\usepackage{amsfonts,amsthm}
\usepackage[caption=false]{subfig}
\usepackage{epsfig,latexsym,graphicx}
\usepackage[numbers, sort, compress]{natbib}
\newtheorem{remark}{Remark}[section]

\begin{document}

\title{\bf{A robust variable screening procedure\\ for ultra-high dimensional data
}}
	
	
	\author{Abhik Ghosh$^1$ and  Magne Thoresen$^2$\\
		$^1$ Indian Statistical Institute, Kolkata, India\\
		$^2$ University of Oslo, Oslo, Norway}
	
	\maketitle

\begin{abstract}
Variable selection in ultra-high dimensional regression problems has become an important issue. 
In such situations, penalized regression models may face computational problems and some pre-screening of the variables may be necessary. 
A number of procedures for such pre-screening has been developed; 
among them the sure independence screening (SIS) enjoys some popularity. 
However, SIS is vulnerable to outliers in the data, and in particular in small samples this may lead to faulty inference. 
In this paper, we develop a new robust screening procedure. 
We build on the density power divergence (DPD) estimation approach and introduce DPD-SIS and its extension iterative DPD-SIS. 
We illustrate the behavior of the methods through extensive simulation studies and 
show that they are superior to both the original SIS and other robust methods when there are outliers in the data. 
We demonstrate the claimed robustness through use of influence functions, and we discuss appropriate choice of the tuning parameter $\alpha$. 
Finally, we illustrate its use on a small dataset from a study on regulation of lipid metabolism.
\end{abstract}

\bigskip\noindent
\textbf{Keywords:} Variable selection; NP dimensionality; Independence screening; 
Minimum density power divergence estimator; Influence Function; Gene selection.
\bigskip

\section{Introduction}
\label{SEC:intro}
	
The introduction of the Omics technologies has led to a revolution in medical research, 
leading to an increased knowledge of the biological background of many diseases and paving the way for personalized therapies. 
A characteristic feature of data arising from the Omics technologies is its high dimensionality, 
which is a challenge for the statistical analysis. A typical example is gene expressions. 
The microarray technology enabled us to measure the expressions of thousands of genes simultaneously, 
while the number of subjects was typically rather low. 
More recently, sequencing technologies allow us to measure genetic features at an even finer resolution, often leading to dimensions in the millions. 
If we are to relate these high-dimensional features to some outcome variable in a regression set-up, 
we need to perform some sort of variable selection 
\citep{Segal/etc:2003,Lusa/etc:2008,Xue/etc:2018,Fu/Wang:2018,Jung/etc:2019,Jacobs/etc:2019,Krzykalla/etc:2020}.

The most commonly used method for identifying important predictor variables in a high-dimensional regression model is to fit a penalized model. 
Consider the linear regression model with response variable $Y$ 
and $p$ explanatory variables (e.g. gene expressions) as covariates,  say $X_1, \ldots, X_p$, given by 
\begin{eqnarray}
Y = \beta_0 + \sum_{j=1}^p \beta_j X_j + \epsilon,
\label{EQ:LRM0}
\end{eqnarray}
where the model error $\epsilon$ is distributed as $\sim N(0, \sigma^2)$.
Given the responses $y_1, \ldots, y_n$ from $n$ independent samples and the corresponding covariate values, say $x_{ij}$, $i=1, \ldots, n$,
for the $j$-th covariate for  $j=1, 2, \ldots, p$, this model can be written in matrix form as
\begin{eqnarray}
y_i = \boldsymbol{x}_i^T\boldsymbol{\beta} + {\epsilon}_i,
\label{EQ:LRM}
\end{eqnarray}
where $\boldsymbol{x}_i=(x_{i1}, \ldots, x_{ip})^T$, 	for $i=1, \ldots, n$.
The model parameters $\boldsymbol{\beta} = (\beta_0, \beta_1, \ldots, \beta_p)^T$ and $\sigma^2$ need to be estimated from the data. 
In the ultra-high dimensional case with $p \gg n$, e.g., Omics data, 
we need to assume sparsity of the regression coefficient $\boldsymbol{\beta}$ to achieve identifiability of the estimators,
i.e., we assume that only a few of the components of $\boldsymbol{\beta}$ are non-zero.
Without loss of generality, we may assume that the true model parameter values are $(\boldsymbol{\beta}_0, \sigma_0^2)$
where $\boldsymbol{\beta}_0^T = (\beta_0, \boldsymbol{\beta}_1, \boldsymbol{0}_{p-s})$ with $\boldsymbol{\beta}_1$ 
being the non-sparse part of size $s \ll n$. Under sparsity assumptions, estimation of the parameters 
$\boldsymbol{\theta}=(\boldsymbol{\beta}, \sigma^2)$ is performed through penalized estimation procedures with appropriate penalties
which can successfully recover all and only the truly important variables (corresponding to non-zero $\beta_j$) asymptotically with 
probability  tending to one. There are plenty of such penalized regression procedures available in recent literature,
starting from the Least Absolute Shrinkage and Selection Operator (LASSO) method of \citet{Tibshirani:1996}
and its refinements \citep[e.g.,][]{Zhang/Huang:2008,Zou:2006} to more advanced procedures based on penalties like SCAD 
\citep{Fan/Li:2001} or MCP \citep{Zhang:2010}, and many more, which work well in moderately high dimensions.
However, a common problem with these methods in ultra-high dimensional set-ups is their computational cost and numerical issues,
which has led to development of simpler variable screening methods at the initial stage to reduce the model size (e.g., number of genetic features)
from the order of millions to an order of a few hundred (often lesser than the sample size as well) 
and then apply an appropriate penalization  method to obtain final model estimates from the reduced set of covariates.
The most popular method for such screening purposes is the Sure Independent Screening (SIS) proposed by \citet{Fan/Lv:2008}
which has a simple interpretation and theoretical guarantees along with fast computation.
Even with its simple structure (the SIS ranks the covariates based on their correlation with the response),
the method yet enjoys the model selection oracle property under ultra-high dimensional set-ups where $\log(p) = O(n^l) $ for some $0<l<1$.
An iterative extension, ISIS, is also proposed in \cite{Fan/Lv:2008} to tackle the issue of collinearity among covariates.
The SIS and ISIS is routinely being applied in ultra-high dimensional applications
and has also been extended to more complex models \citep[e.g.,][]{Fan/Song:2010,Barut/etc:2016,Zhao/Li:2012,Luo/etc:2014}.
However, one major drawback of the SIS or ISIS is its non-robust nature against data contamination as indicated 
already in the discussion of the original paper itself. This issue can be crucial when applying the method
for screening of important genes from large scale Omics data, which are often prone to at least a few outliers.
Here is our motivating data example.

\bigskip
\noindent
\textbf{Triglyceride Data:}\\
	Intake of marine omega-3 fatty acids may reduce the risk of cardiovascular disease (CVD), especially in high-risk individuals. 
	Elevated serum triglyceride (TG) levels are strongly associated with increased risk of CVD, 
	and the CVD risk reducing effect of marine omega-3 fatty acids is thought to be mainly mediated through reduction of serum TG levels.
	However, it is well known that there is large individual variation with regard to TG response in relation to intake of fatty acids, 
	and an improved understanding of such individual variation would be beneficial.
	We are analyzing data from a small randomized study ($n=54$) where the subjects received either fish oil, 
	oxidized fish oil or sunflower oil for a period of seven weeks, and serum TG levels were measured at baseline and after seven weeks.
	Our goal is to relate TG response (the difference between two measurements of serum TG levels) to gene expressions measured at baseline. 
	Gene expressions were measured using microarray technology, and we have available data from in total $p=21236$ probes.
	Thus, an initial variable screening to reduce the model size is needed.

	\begin{figure}[!b]
		\centering
		\subfloat[Boxplot of TG Difference]{
			\includegraphics[width=0.4\textwidth]{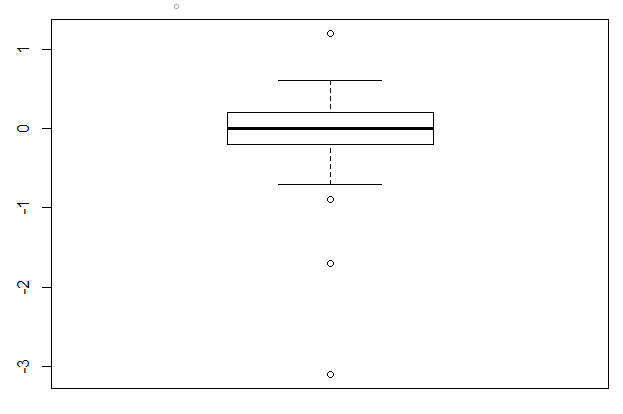}
			\label{FIG:boxplot_Y}}
		\\	
		\subfloat[Gene: FOXF2]{
			\includegraphics[width=0.45\textwidth]{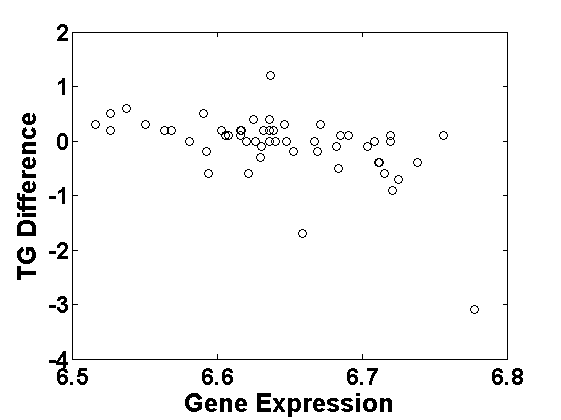}
			\label{FIG:scPlot1}}
		\subfloat[Gene: MORC4]{
			\includegraphics[width=0.45\textwidth]{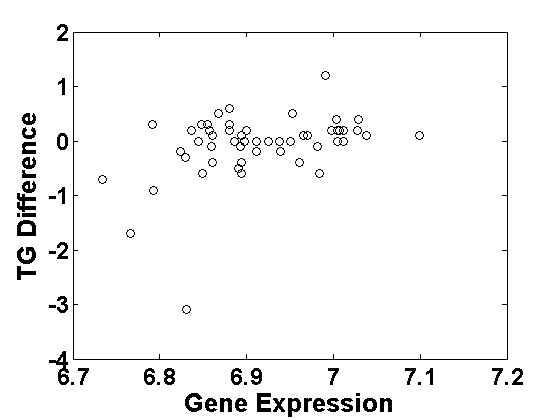}
			\label{FIG:scPlot2}}
		\\
		\subfloat[Gene: EEF1A1]{
			\includegraphics[width=0.45\textwidth]{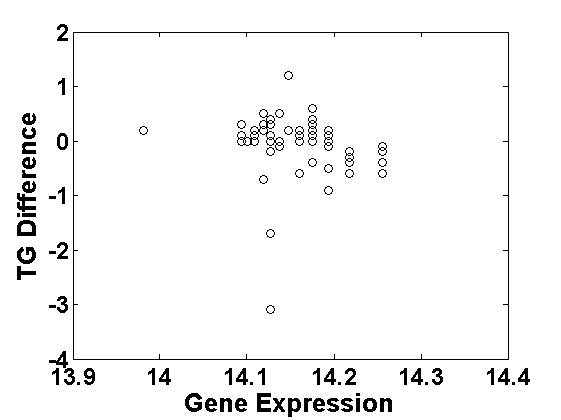}
			\label{FIG:scPlot3}}
		\subfloat[Gene: ZSCAN12]{
			\includegraphics[width=0.45\textwidth]{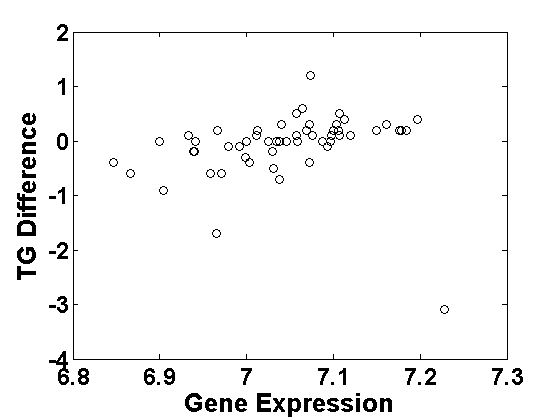}
			\label{FIG:scPlot4}}
		\caption{Box-plot of Response and Scatter plots of the response against different Genes}
		\label{FIG:MSE_n100}
	\end{figure}

	From the box-plot of the TG response (Fig \ref{FIG:boxplot_Y}), 
	it is clearly justified to fit a normal error distribution in the linear regression model (\ref{EQ:LRM})
	except for three outlying values. Now, if we are screening the genes via correlation with response in SIS or ISIS,
	these outliers will have an erroneous effects. Note that, it is not justified to remove these outliers at the start of the screening procedure,
	since they are outliers in the univariate distribution of response only but 
	may or may not be outliers in the bivariate distribution of the response with any covariate.
	A few such examples are given in Fig \ref{FIG:scPlot1}--\ref{FIG:scPlot4}; 
	the outliers seem more legitimate in the bivariate space for the first case (Figure \ref{FIG:scPlot1}),
	having no effect on the significance of the associated regression slope,
	but this is not true for the other cases.  
	In Figure \ref{FIG:scPlot2}, the outliers make the relation between the response $Y$ and the corresponding gene look significant 
	and hence it will come up towards the top of the selected gene list through usual SIS,
	although there is clearly no relation among these variables after removing the outliers.
	The situation is more serious in the last two cases (Figures \ref{FIG:scPlot3}-\ref{FIG:scPlot4}); 
	there are actually strong associations between the response variable and both these genes 
	which get masked by the presence of outliers and hence, these genes will not be among the top selected genes  in SIS or even through ISIS 
	(we have checked up to three steps of ISIS). 
	Further, Figure \ref{FIG:scPlot3} also presents a new outlier in the covariate space (in the gene expressions)
	and the same outlier may be present in several other gene expressions as well;
	such a scrutiny for each gene is clearly not feasible with larger sets of Omics data with potentially millions of features.
	Even if performed (with a huge time effort), this might leave us with very few cases left for performing any reasonable (joint) inference.
	A robust screening method which would ignore the effect of such outliers would be of great help in such ultra-high dimensional problems.
	\hfill{$\square$}

	\bigskip
	In this paper, we will develop a new robust screening procedure, an extension of the usual SIS, 
	using the popular density power divergence (DPD) based estimation approach (briefly described in Section \ref{SEC:MDPDE}). 
	The DPD measure was originally proposed by \citet{Basu/etc:1998} in the context of robust estimation in IID data.
	It has recently become very popular for robust inference in general and is widely applied on different types of data; 
	see, e.g., \cite{Basu/etc:2011}.
	The same approach has also been used for high-dimensional penalized linear regression and variable selection more recently 
	\citep{Zang/etc:2017,Ghosh/Majumdar:2019} and has been shown to be extremely useful under data contamination along with 
	consistency of the resulting estimates and oracle model selection property.
	However, the computation is still a concern in ultra-high dimensional set-ups 
	and a robust version of SIS along the same line would be a useful approach to analyze such data more robustly,
	with robust screening at an initial stage followed by the robust DPD based penalized regression method to the reduced low or moderately high dimensional set of covariates.
	In the current work we fill the gap in the literature for the first (screening) stage by proposing a robust screening method based on the DPD
	for ultra-high dimensional linear regression models and illustrate its claimed robustness property theoretically as well as numerically.
	A robust version of ISIS along the same line using DPD will also be discussed to tackle the correlations among covariates. 
	The suggested method will be applied to our motivating data example.

	We also compare our method with the existing state-of-the-art robust screening procedures for ultra-high dimensional linear models
	\citep[etc.]{Hall/Miller:2009,Li/etc:2012a,Li/etc:2012b,Mu/Xiong:2014,Wang/etc:2017}
	through extensive simulation studies.
	The major advantages of our proposed DPD based SIS and ISIS methods can be summarized as follows.
	\begin{itemize}
		\item Most (if not all) of the existing robust screening procedures are non-parametric in nature.
		We consider a more efficient parametric approach with robust parameter estimation  
		to develop our robust sure screening procedure (in Section \ref{SEC:DPD_Screening}).
		Thus, when the assumed (parametric) linear regression model is even approximately correct,
		our proposed screening procedure enjoys a significantly improved performance over the existing non-parametric robust versions of SIS
		(as evident from simulations in Section \ref{SEC:simulation}).  
		
		\item As a consequence of the parametric approach, 
		our proposed screening method indeed estimates the error variance ($\sigma^2$) also from data in each step 
		(see Sections \ref{SEC:MDPDE}-\ref{SEC:DPD_SIS}), 
		rather than just assuming it to be known (not possible in practice) or ignoring it as in most existing (non-parametric) approaches.
		This, in turn, provides better control 
		of the signal-to-noise ratio in each step of the screening, eventually leading to improved variable selection behavior.
		
		\item Through extensive simulation studies, we have illustrated the performance of the proposed DPD based screening procedures
		with suitably chosen tuning parameter value ($\alpha$). Under pure data, as expected with any robust procedure, 
		the performance of the proposed screening procedures become slightly inferior compared to the usual SIS;
		however, the loss is small for smaller values of the underlying tuning parameter $\alpha$ and additionally, 
		they significantly outperform all the existing (non-parametric) robust procedures.
		Under contamination, our proposal performs the best, selecting the most stable set of variables even 
		when the contamination is as heavy as 20\%, whereas the usual SIS breaks down even in the presence of 5\% contamination.
		A few existing non-parametric procedures like those based on ranks or G-K correlation \citep{Gnanadesikan/Kettenring:1972}
		provide robust results under contaminations which are competitive to our proposal, but our proposed screening procedure
		even outperform them under more vulnerable cases like  heavier contaminations, weak signal-to-noise ratios or smaller sample sizes.
		This makes our proposal even more advantageous over all the existing variable screening procedures. 
	\end{itemize}
	
In continuation with the last bullet point above, we would like to emphasize the size of our empirical experiments
where we have considered $p=5000$ covariates with a sample size as small as $n=50$. 
Such an experimental set-up is more extreme compared to most (if not all) other existing work on SIS or its extensions 
for the linear regression model. 
However, they are clearly more realistic scenarios arising in medical sciences  and hence provide more confidence about 
the performance of the proposed method in real data applications.

\section{Background: The Minimum DPD Estimators} 
\label{SEC:MDPDE}

	For completeness, we start with a brief description of the DPD measure and associated minimum divergence estimation process.
	Given two densities $f_1, f_2$ with respect to a common dominating measure $\mu$, the DPD between them is defined, 
	in terms of a tuning parameter $\alpha \geq 0$, as \citep{Basu/etc:1998}
	
	\begin{eqnarray}
	d_\alpha(f_1,f_2) &=& \int  f_2^{1+\alpha} - \frac{1+\alpha}{\alpha} \int f_2^\alpha f_1 
	+ \frac{1}{\alpha} \int f_1^{1+\alpha}, ~~~~~\mbox{ if } ~~ \alpha > 0, \nonumber \\
	\mbox{and }~~~~~
	d_0(f_1,f_2) &=& \lim_{\alpha \rightarrow 0} d_\alpha(f_1,f_2) = \int f_1 \log(f_1/f_2). 
	\nonumber
	\end{eqnarray}
	Given IID samples $y_1, \ldots, y_n$ from a true density $g$ to be modeled by a parametric family of densities
	$\mathcal{F}=\{f_{\boldsymbol{\theta}}: \boldsymbol{\theta}\in \theta \}$,
	the minimum DPD estimator (MDPDE) of $\boldsymbol{\theta}$ is obtained as a minimizer of 
	$d_\alpha(\widehat{g}, f_{\boldsymbol{\theta}})$ with respect to $\boldsymbol{\theta}\in\theta$,
	where $\widehat{g}$ is an empirical estimate of the true density $g$.
	One of the reasons for the popularity of the DPD measure is that, unlike many other divergences,
	we can avoid the non-parametric density estimation (e.g., kernel for continuous models) in this estimation step.
	To see this, note that the third integral in $d_\alpha(g, f_{\boldsymbol{\theta}}) $ is independent of $\boldsymbol{\theta}$,
	so we only need to empirically estimate the second integral $\int f_{\boldsymbol{\theta}}^\alpha gd\mu = \int f_{\boldsymbol{\theta}}^\alpha dG$
	where $G$ is the true distribution function. Therefore, we can just use the empirical distribution function to estimate $G$
	to obtain the estimates of the second integral as $n^{-1}\sum_{i=1}^n f_{\boldsymbol{\theta}}^\alpha(y_i)$.
	Hence, avoiding the complications of non-parametric smoothing, 
	the MDPDE can be obtained by minimizing a simpler objective function, namely
	$$
	H_{n,\alpha}(\boldsymbol{\theta}) = \int f_{\boldsymbol{\theta}}(y)^{1+\alpha}d\mu(y) 
	- \frac{1+\alpha}{\alpha}\frac{1}{n}\sum_{i=1}^n f_{\boldsymbol{\theta}}^\alpha(y_i) + \frac{1}{\alpha}.
	$$
	The last term, $1/\alpha$, is added to ensure that 
	$$
	\lim_{\alpha \rightarrow 0} H_{n,\alpha}(\boldsymbol{\theta}) = - 
	\frac{1}{n}\sum_{i=1}^n \log f_{\boldsymbol{\theta}}(y_i),
	$$ 
	so that the MDPDE at $\alpha=0$ coincides with the classical (but non-robust) MLE.
	It has been observed that the robustness of these MDPDEs increases with increasing $\alpha$ 
	but their asymptotic efficiency decreases slightly, although this loss in efficiency 
	is not significant for smaller $\alpha>0$; see \cite{Basu/etc:2011} for more detailed theory and examples.
	Due to the simple interpretation as a robust generalization of the MLE,
	high robustness along with high efficiency and easy computation, the MDPDE has become an extremely popular estimator in parametric robust inference 
	and has been extended to different types of more complex datasets;
	see, e.g., \cite{Ghosh/Basu:2013,Ghosh/Basu:2015a,Ghosh/Basu:2016,Ghosh/Basu:2019,Ghosh:2019,Ghosh/etc:2018} among many others.

	In particular, \citet{Durio/Isaia:2011} discussed the MDPDE for linear regression models
	and its theoretical properties are later established by \citet{Ghosh/Basu:2013}
	in the context of more general independent non-homogeneous set-ups.
	In this set-up, we assume that the sample $y_i$s ($i=1, \ldots,n$) are independent 
	but potentially have different densities $g_i$ which are then modeled by the family of densities
	$\mathcal{F}_i=\{f_{i,\boldsymbol{\theta}}: \boldsymbol{\theta}\in \theta \}$, respectively. 
	The most common example is the parametric regression model, where the distribution of the response $y_i$ 
	depends of the given ($i$-th) value of the covariates but they all share the same parameter sets 
	-- the regression coefficients (and error variance).
	For such set-up, \citet{Ghosh/Basu:2013} proposed to obtain the MDPDE of $\boldsymbol{\theta}$ 
	by minimizing the average of the DPD measures $d_\alpha(\widehat{g}_i, f_{i,\boldsymbol{\theta}})$ 
	over different possible densities ($i=1, \ldots, n$) with respect to $\boldsymbol{\theta}\in\theta$,
	which leads to the following simpler objective function (along the same lines as in the IID case)
	\begin{eqnarray}
	H_{n,\alpha}(\boldsymbol{\theta}) = \int f_{\boldsymbol{\theta}}(y)^{1+\alpha}d\mu(y) 
	- \frac{1+\alpha}{\alpha}\frac{1}{n}\sum_{i=1}^n f_{\boldsymbol{\theta}}^\alpha(y_i) + \frac{1}{\alpha}.
	\label{EQ:DPD_ObjFunc_INH}
	\end{eqnarray}
	Here, also the MDPDE at $\alpha=0$ coincides with the corresponding MLE.
	Theoretical properties of the MDPDEs under this general non-homogeneous set-up is derived in \cite{Ghosh/Basu:2013},
	which are then applied to several types of regression models  \cite{Ghosh/Basu:2016,Ghosh/Basu:2019,Ghosh:2019}.

\section{Proposed Robust Variable Screening Procedures}
\label{SEC:DPD_Screening}

\subsection{The DPD-SIS}
\label{SEC:DPD_SIS}
	
We now consider the linear regression model (\ref{EQ:LRM}) with ultra-high dimensional covariates
and the true sparse regression coefficient $\boldsymbol{\beta}_0$ as described in Section \ref{SEC:intro};
let us denote the true sparse model as $\mathcal{M}_0 = \left\{ 1\leq j \leq p : \beta_{0j} \neq 0 \right\} = \{1, 2, \ldots, s\}$.
Recall that the SIS method \citep{Fan/Lv:2008} can also be considered as ordering the absolute value of the slope in 
marginal regression models of the response with individual (standardized) covariates.
Given values of  the $j$-th covariate $X_j$ for each $j=1, \ldots, p$,
we consider the $j$-th marginal model
\begin{eqnarray}
y_i = \gamma_j + \beta_j x_{ij} + \epsilon_{ij}, ~~~~i=1, \ldots,n,
\label{EQ:LRM_Marginal}
\end{eqnarray}
where the $\epsilon_{ij}$s are IID for $i=1, \ldots, n$, each having distribution $N(0, \sigma_j^2)$.
We estimate the parameters $\boldsymbol{\theta}_j=(\gamma_j, \beta_j, \sigma_j)^T$ by usual MLE or OLS based methods,
say, $(\widehat{\gamma}_j, \widehat{\beta}_j, \widehat{\sigma})$.
Note that, when all covariates are standardized, ranking them in order of (absolute) correlation with the response
is equivalent to ordering the estimated slopes $|\widehat{\beta}_j|$. However, this method is clearly non-robust 
since the estimates  $\widehat{\beta}_j$s are so for MLE/OLS.

Here, we will propose to use the same approach as in the usual SIS, 
but with robust estimates for $\beta_j$ in the marginal model using the DPD approach. 
Let us fix a $j\in\{1, 2, \ldots, p \}$ and an $\alpha>0$. 
Since, given covariate values, $y_i \sim N(\gamma_j+ \beta_jx_{ij}, \sigma_j^2)$, 
it belongs to the non-homogeneous set-up discussed in Section \ref{SEC:MDPDE}
and hence, we can define the MDPDE of the parameters $\boldsymbol{\theta}_j$ via the objective function (\ref{EQ:DPD_ObjFunc_INH}) there.
For the marginal model (\ref{EQ:LRM_Marginal}), 
one can easily simplify the MDPDE objective function as to have the form
$H_{n,\alpha}(\boldsymbol{\theta}_j) = \frac{1}{n}\sum\limits_{i=1}^n l_\alpha\left(y_i, \gamma_j + \beta_j x_{ij}, \sigma\right)$,
where
	\begin{eqnarray}
	l_\alpha\left(y, \eta, \sigma\right) = \frac{1}{\sigma^{\alpha}(2\pi)^{\alpha/2}}
	\left(\frac{1}{\sqrt{1+\alpha}} - \frac{1+\alpha}{\alpha} e^{-\frac{\alpha(y - \eta)^2}{\sigma^2}}\right) 
	+ \frac{1}{\alpha},  
	\label{EQ:DPD_lossGen_Normal}
	\end{eqnarray}
	Then, we define the MDPDE of $\boldsymbol{\theta}_j$ for the marginal model (\ref{EQ:LRM_Marginal}) as 
	\begin{eqnarray}
	\widehat{\boldsymbol{\theta}}_j^M = (\widehat{\gamma}_j^{M\alpha}, \widehat{\beta}_j^{M\alpha}, \widehat{\sigma}_j^{M\alpha})
	=\arg\min\limits_{\boldsymbol{\theta}} \frac{1}{n}\sum_{i=1}^n l_\alpha\left(y_i, \gamma_j + \beta_j x_{ij}, \sigma\right).
	\label{EQ:MMDPDE}
	\end{eqnarray}
	This is a much simpler optimization problem with only three parameters (compared to any penalized estimation problem),
	but we need to run it $p$ times (once for each $j=1, \ldots, p$).
	However, the overall computation time is still much lower than the penalized regression procedure with NP dimensional $p$.
	Based on these MDPDEs for a given $\alpha>0$,  we can now choose the important variables 
	in order of the values of $|\widehat{\beta}_j^{M\alpha}|$, which we refer to as the DPD-SIS procedure;
	for given index $d$ we select the estimated model $\widehat{\mathcal{M}}_\alpha(d)$.
	Once $\widehat{\mathcal{M}}_\alpha(d)$ is obtained, 
	one can then apply any suitable penalized regression method 
	on the reduced set of covariates from $\widehat{\mathcal{M}}_\alpha(d)$ 
	to obtain the final spares estimate of the parameters in the original model (\ref{EQ:LRM}).
	For this stage, to be consistent with the DPD-SIS, 
	we suggest to use the penalized DPD based estimation procedure with the same $\alpha>0$ as used in the screening stage,
	following \citet{Ghosh/Majumdar:2019}. 
	Therefore, our proposed robust screening procedure, the DPD-SIS, can be summarized in the following algorithm.\\
	
	\noindent\textbf{Algorithm 1: DPD-SIS($\alpha$)}  
	\begin{enumerate}
		\item \textbf{Input:} $n$-vector of responses $\boldsymbol{y}$; $n\times p$ matrix of covariates $\boldsymbol{X}$; screening model size $d$.
		\item For each $j=1, \ldots, p$, compute the marginal MDPDE $\widehat{\beta}_j^{M\alpha}$ via (\ref{EQ:MMDPDE}).\\
		(This is an optimization in three parameters only and can be performed either by a standard optimization function in some software or 
		by standard numerical techniques).
		\item Sort $|\widehat{\beta}_j^{M\alpha}|$ in decreasing order for $j=1, \ldots, p$.\\ 
		Set $r_k = j$, if $|\widehat{\beta}_j^{M\alpha}|$ has rank $k$, for $k=1, \ldots, p$.
		
		\item Construct the estimated model set $\widehat{\mathcal{M}}_\alpha(d) = \{ r_1, \ldots, r_d \}$, 
		with indices corresponding to the top $d$ values of (absolute) marginal MDPDEs. 
		
		\item Run a robust penalized regression model (low or moderate dimensional) with the covariates selected in $\widehat{\mathcal{M}}_\alpha(d)$
		to obtain an estimated coefficient vector, 
		say $\widehat{\boldsymbol{\beta}}_d =(\widehat{\beta}_{d0}, \widehat{\beta}_{dr_1}, \ldots, \widehat{\beta}_{dr_d})^T$.\\
		(We suggest to use the DPD based method of \citet{Ghosh/Majumdar:2019} with the same $\alpha$, 
		which also gives an estimate $\widehat{\sigma}^2$ of the overall model error variance $\sigma^2$.)
		
		\item \textbf{Output:} The final estimated model 
		$\widehat{\mathcal{M}} =  \left\{ 1\leq k \leq d : \widehat{\beta}_{dr_k} \neq 0 \right\}$ 
		along with the parameter estimates  $\widehat{\boldsymbol{\beta}}_d$ (and the estimate $\widehat{\sigma}^2$ of $\sigma^2$, if available).	 
	\end{enumerate}

	Note that, at $\alpha=0$ (in a limiting sense), the marginal MDPDE of regression coefficients just coincides 
	with the MLE and hence with the OLS as well.
	Thus the proposed DPD-SIS algorithm at $\alpha=0$ becomes close to the usual SIS of \citet{Fan/Lv:2008},
	but differ from it in that we are also estimating $\sigma$ in DPD-SIS (even at $\alpha=0$) rather than estimating only $\beta_j$ in 
	the usual SIS. This makes DPD-SIS at $\alpha=0$ to have slightly better performance than the usual SIS
	even under no data contamination (pure data) due to better control based on the signal-to-noise ratio in each step;
	these claims are supported by our simulations presented in Section \ref{SEC:simulation}
	where this method is referred to as {SIS-Reg}.
	This improvement due to the estimation of the error variance in each step is also present in the DPD-SIS with any $\alpha>0$,
	another advantage of our proposed robust screening procedure over the existing ones.
	The extent of robustness of these DPD-SIS procedures further increases with increasing values of $\alpha>0$, 
	as will be justified and illustrated further in Sections \ref{SEC:robust} and \ref{SEC:simulation}.

\subsection{Iterative DPD-SIS}
	\label{SEC:DPD_ISIS}
	
	It has been noted that the usual SIS fails to pick up a variable having weak marginal correlation but significant joint relation with the response; 
	on the other hand, it might pick up a variable having stronger marginal correlation but no joint relation with the response. Such cases occur
	mostly due to strong correlation between the important and unimportant predictor variables. To solve these issues,
	\citet{Fan/Lv:2008} also proposed an iterative extension of SIS, namely the ISIS, 
	which selects the truly important variables even under the above situations.
	Later, several extensions of the original ISIS have also been proposed \citep{Saldana/Feng:2018}.
	As a robust extension of SIS, the DPD-SIS also suffers from the above issues, and fails to provide optimal results
	when covariates are strongly correlated (see Section \ref{SEC:simulation}) and an iterative extension in the line of ISIS is required. 
	
	The DPD-SIS can be easily extended through iterations to avoid the strong effects of correlation among predictors
	by considering, in subsequent iterations, the residuals from the fitted regression with predictors picked up in earlier stages.
	More explicitly, we start with DPD-SIS (Algorithm 1) in the first step to select $k_1$ variables with index set 
	$\mathcal{A}_1 = \{i_1, \ldots, i_{k_1}\}$. Then, in the second step, we compute the residuals from 
	the fitted regression model of the response $\boldsymbol{y}$ on the selected covariates in $\mathcal{A}_1$.
	The DPD-SIS screening is again applied taking these residuals as our new response 
	to select another $k_2$ variables from the pool of variables with index set $\{1, 2, \ldots, p \} \setminus \mathcal{A}_1$;
	let us denote the index set of these $k_2$ selected variables as $\mathcal{A}_2$.
	We further proceed repeating these steps to generate the index sets $\mathcal{A}_3, \ldots, \mathcal{A}_l$ 
	of selected variables in the subsequent stages till we reach our target model size, say $d$, i.e.,
	till the smallest $l$ for which $|\cup_{i=1}^l \mathcal{A}_i|= d$.
	Considering its similarity with the ISIS, we refer to this robust iterative variable screening procedure as 
	Iterative DPD-SIS or, in short, DPD-ISIS,  which is presented schematically in the following algorithm. 
	
	\bigskip
	\noindent\textbf{Algorithm 2: DPD-ISIS($\alpha$)}
	\begin{enumerate}
		\item \textbf{Input:} $n$-vector of responses $\boldsymbol{y}$; $n\times p$ matrix of covariates $\boldsymbol{X}$; screening model size $d$.
		
		\item Set $i=1$, $\boldsymbol{y}^{(1)}=\boldsymbol{y}$ and index set of available covariates as $\mathcal{W}_1=\{1, \ldots, p\}$ 
		
		\item \textbf{DPD-SIS with model size $d'$:}
		\begin{enumerate}
			\item For each $j\in \mathcal{W}_1$, compute the marginal MDPDE $\widehat{\beta}_j^{M\alpha}$ via (\ref{EQ:MMDPDE}) with response $\boldsymbol{y}^{(i)}$
			and covariate $X_j$.
			\item Sort $|\widehat{\beta}_j^{M\alpha}|$ in decreasing order for $j\in \mathcal{W}_i$ and set $r_k = j$, if $|\widehat{\beta}_j^{M\alpha}|$ has rank $k$.
			
			\item Construct the estimated model set $\widehat{\mathcal{M}}_\alpha^{(i)} = \{ r_1, \ldots, r_{d'} \}$, 
			with indices corresponding to the top $d'$ values of (absolute) marginal MDPDEs.
		\end{enumerate}
		
		\item Run any suitable (fast) robust penalized regression model (e.g., RLARS \citep{Khan/etc:2007}) 
		with the main response $\boldsymbol{y}$ and the covariates selected in 
		$\cup_{k=1}^i  \widehat{\mathcal{M}}_\alpha^{(k)}$ 
		to get estimated coefficient vector $\widehat{\boldsymbol{\beta}}^{(i)}$.\\
		Let us assume that, at this $i$-th stage, the number of covariates selected in $\cup_{k=1}^i  \widehat{\mathcal{M}}_\alpha^{(k)}$ is $k_i$ and denote them as $\{ j_1, \ldots, j_{k_i}\}$  
		so that the estimated coefficient vector has the form $\widehat{\boldsymbol{\beta}}^{(i)} =(\widehat{\beta}_{0}^{(i)}, \widehat{\beta}_{j_1}^{(i)}, \ldots, \widehat{\beta}_{j_{k_i}}^{(i)})^T$.
		Denote $\mathcal{A}_i = \left\{ j_a : \widehat{\beta}_{j_a}^{(i)} \neq 0, a=1, \ldots, k_i \right\} \subset \mathcal{W}_1$. 
		
		\item If a specified stopping criterion (see discussion below) is satisfied, go to step 8.  Otherwise go to Step 6.
		
		\item Compute the residuals $\boldsymbol{r}^{(i)} = \boldsymbol{y} - \boldsymbol{X}_{\mathcal{A}_i}\widehat{\boldsymbol{\beta}}^{(i)}$.
		
		\item Set $\boldsymbol{y}^{(i+1)}=\boldsymbol{r}^{(i)}$ and the index set of available covariates as $\mathcal{W}_i=\mathcal{W}_1 \setminus\mathcal{A}_i$.\\
		Change $i$ to $i+1$ and go to Step 3. 
		
		\item Run a robust penalized regression model (low or moderate dimensional) 
		with the covariates selected in $\mathcal{A}_i$ to get estimated coefficient vector, 
		say $\widehat{\boldsymbol{\beta}}_d =(\widehat{\beta}_{d0}, \widehat{\beta}_{dr_1}, \ldots, \widehat{\beta}_{dr_d})^T$.
		
		\item \textbf{Output:} The final estimated model 
		$\widehat{\mathcal{M}} =  \left\{ 1\leq k \leq d : \widehat{\beta}_{dr_k} \neq 0 \right\}$ 
		along with the parameter estimates  $\widehat{\boldsymbol{\beta}}_d$ (and the estimate $\widehat{\sigma}^2$ of $\sigma^2$, if available).	 
	\end{enumerate} 
	
	A few remarks related to the above algorithm is in order before further discussions. 
	Firstly, the most straightforward stopping criterion (required in Step 5) could be $|\mathcal{A}_i|<d$. 
	Step 8 assumes that the size of $\mathcal{A}_i$, at the end of the last iteration, is exactly $d$, 
	which may not always be the case.
	When $|\mathcal{A}_i|>d$, we may work with all those selected variables or remove the extra variables having lower 
	values of the marginal MDPDEs at the last stage of iteration. 
	Alternatively the DPD-ISIS can also be terminated after a pre-fixed number of iterations (say $i=i_{\max}$)
	or when the size of the active set does not change from its value in the previous iteration 
	$(i.e., |\mathcal{A}_i| = |\mathcal{A}_{i-1}|)$. 
	
	Secondly, in step 4 of DPD-ISIS, 
	any fast robust  penalized regression  method, like RLARS, may be used without hampering the basic structure of DPD-ISIS. 
	However, we strongly suggest to use the DPD based penalized regression method of \citet{Ghosh/Majumdar:2019} with the same $\alpha$
	in Step 8 to obtain the final model; as in DPD-SIS, it makes the whole procedure structurally consistent and also 
	provides an estimate $\widehat{\sigma}^2$ of the overall error (unexplained) variance $\sigma^2$ in our final model.
	
	Finally, it is worthwhile to note that our algorithm of DPD-ISIS is more similar to an extension of ISIS, 
	namely Van-ISIS described in \cite{Saldana/Feng:2018}, rather than its original version proposed in \cite{Fan/Lv:2008}.
	The difference is mainly in Step 4 of the algorithm, where we consider all the covariates selected till the $i$-th iteration 
	in the penalized joint regression model in the $i$-th stage, as in Van-ISIS; 
	hence a variable which has been selected in an earlier stage could have been removed at the $i$-th stage due to insertion of new variables in the model.
	The original version of \citet{Fan/Lv:2008}  considered the penalized regression to be run with only the variables selected in that $i$-th iteration 
	(and not the previously selected covariates) and hence a false positive selected at one iteration cannot be removed at any subsequent iteration.
	In this method the model size continue to increase whereas, in our approach, it may grow or shrink depending on 
	the joint relationship of all the variables selected, reducing the number of false-positive covariates.
	
	Although the DPD-SIS at $\alpha=0$ is similar to the usual (non-robust) regression based SIS,
	the DPD-ISIS($\alpha=0$) is slightly different from its usual non-robust counterpart van-ISIS.
	Due to the use of RLARS within iterations, the DPD-ISIS($\alpha=0$) is slightly more robust than van-ISIS
	and additionally DPD-ISIS at any $\alpha$ (including 0) estimates the error variance in a marginal regression setting
	whereas the usual van-ISIS uses marginal correlation based screening.
	However,  the DPD-ISIS at $\alpha=0$ is not yet acceptable as a robust method 
	since the estimates of the marginal regression coefficients are still non-robust (MLE).
	As $\alpha>0$ increases, the proposed DPD-ISIS becomes more robust, which is justified 
	theoretically through the influence function of the underlying marginal MDPDE  in the following section.

	\section{Robustness of the proposed DPD-SIS procedure}
	\label{SEC:robust}

	The proposed variable screening methods DPD-SIS and DPD-ISIS both depend crucially on the estimates $\widehat{\boldsymbol{\beta}}_j^M$
	from each marginal regression model and hence, the same is true for their robustness. 
	If these marginal estimates are robust with respect to any outliers or noise contamination in either the response or the corresponding covariate,
	their ordering (in absolute value) is also expected to be robust under data contamination, leading to correct and stable variable screening.

\begin{figure}[!h]
		\centering
		\subfloat[$\alpha=0.1$]{
			\includegraphics[width=0.4\textwidth]{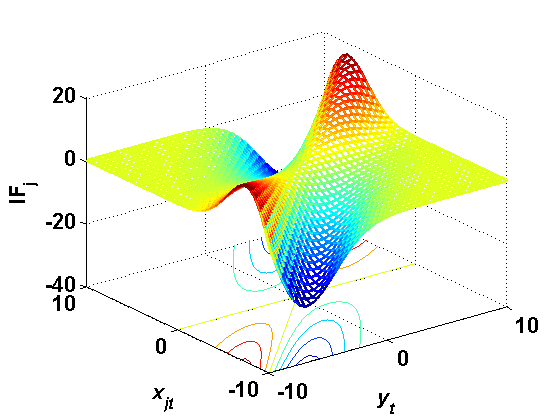}
			\label{FIG:IF01}}
		\subfloat[$\alpha=0.3$]{
			\includegraphics[width=0.4\textwidth]{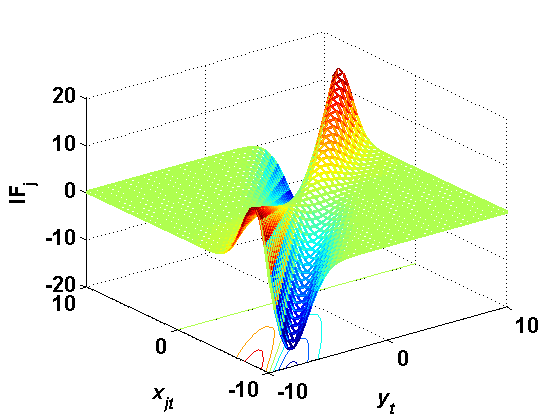}
			\label{FIG:IF03}}
		\\
		\subfloat[$\alpha=0.5$]{
			\includegraphics[width=0.4\textwidth]{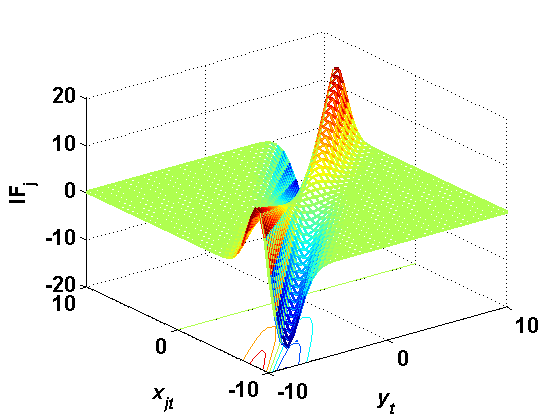}
			\label{FIG:IF05}}
		\subfloat[$\alpha=1$]{
			\includegraphics[width=0.4\textwidth]{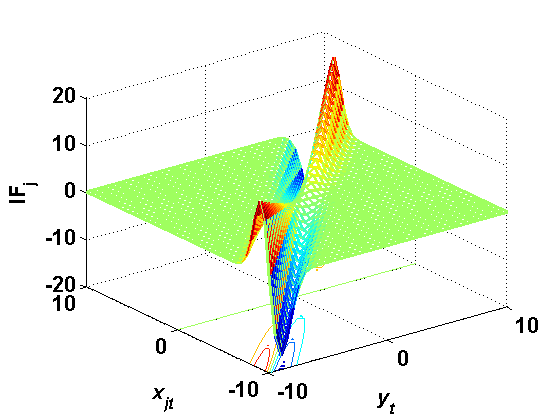}
			\label{FIG:IF10}}
		\caption{Influence functions ($\mathcal{IF}_j$) of the marginal MDPDEs for different values of $\alpha>0$}
		\label{FIG:IF}
	\end{figure}

	The robustness of the MDPDE under any parametric set-up, including the linear regression model, is well-studied in the literature
	\citep{Basu/etc:2011,Ghosh/Basu:2013,Ghosh/Basu:2016}; it is known to crucially depend on the choice of $\alpha$.
	At $\alpha=0$ the MDPDE coincides with the MLE and hence, it is extremely non-robust; its robustness increases with increasing values of $\alpha>0$.
	This can be examined theoretically through the concept of influence functions and gross error sensitivity \citep{Hampel/etc:1986}. 
	The influence function indeed measures the asymptotic (standardized) bias of the estimator caused by an infinitesimal contamination 
	through the degenerate distribution at a distant outlier point.
	For our $j$-th marginal regression model (\ref{EQ:LRM_Marginal}), the influence function of the MDPDE estimator 
	$\widehat{\beta}_j^{M\alpha}$ of the regression coefficients $\beta_j$
	with respect to the contamination point,  say $y_t$,  in the response for a given covariate value, say $x_{jt}$, 
	can be obtained from the general results of \citet{Ghosh/Basu:2013}. 
	When the assumed linear model is true with parameter values  
	$(\gamma_j^{(0)}, \beta_{j}^{(0)}, \sigma_j^{(0)})$, 
	it has a simplified form given by 	
	\begin{eqnarray}
	\mathcal{IF}_j^{(\alpha)}(y_t|x_{jt}) &=&(1+\alpha)^{3/2} \frac{(x_{jt}-E(X_j))}{Var(X_j)}\left(y_t - \gamma_j^{(0)} - \beta_j^{(0)}x_{jt}\right)
	e^{-\frac{\alpha\left(y_t - \gamma_j^{(0)} - \beta_j^{(0)}x_{jt}\right)^2}{2(\sigma_j^{(0)})^2}}.
	\nonumber
	\end{eqnarray}
	\\
	To study its nature, in Figure \ref{FIG:IF} we plot $\mathcal{IF}_j^{(\alpha)}(y_t| x_{jt})$ over the contamination point $y_t$ 
	for different values of $\alpha>0$, by taking the true parameter values to be $(\gamma_j^{(0)}, \beta_{j}^{(0)}, \sigma_j^{(0)})=(0,1,1)$.
	We assume $E(X_j)=0$ and $Var(X_j)=1$. 
	For the case $\alpha=0$, the above influence function simplifies to a linear function of $y_t$,
	and hence, it is unbounded with respect to the contamination point $y_t$, for all possible covariate values $x_{jt}$
	which justifies the well-known non-robust nature of the MLE (MDPDE at $\alpha=0$). 
	However, it can be clearly seen from Figure \ref{FIG:IF} that, for the proposed DPD-SIS at any $\alpha>0$, 
	the influence function of the corresponding marginal estimator $\widehat{\beta}_j^{M\alpha}$
	is bounded in the contamination point $y_t$ for all values of $x_t$,
	indicating the claimed robust nature.

	\begin{figure}[!h]
		\centering
		\subfloat[Sensitivity for different values of $\delta$]{
			\includegraphics[width=0.47\textwidth]{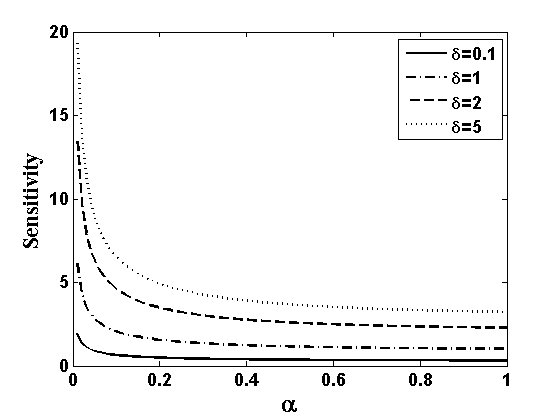}
			\label{FIG:Sensitivity}}
		\subfloat[Asymptotic relative efficiency(ARE)]{
			\includegraphics[width=0.47\textwidth]{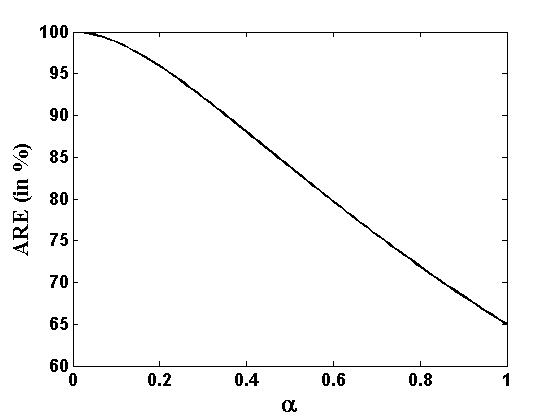}
			\label{FIG:ARE}}
		\caption{Illustration of the trade-off between the robustness and efficiency of the marginal MDPDEs over $\alpha$}
		\label{FIG:Trade-off}
	\end{figure}
	
	Further, we also study their self-standardized gross error sensitivity,  
	which is the maximum of the $L_2$-norm of the influence functions, 
	standardized by the variance of the MDPDE, over all possible contamination points.
	It is clearly observed from Figure \ref{FIG:Sensitivity} that
	these sensitivity measures indeed decrease with increasing  $\alpha>0$ 
	for any given value of $\delta={(x_{jt}-E(X_j))^2}/{Var(X_j)}$; 
	thus the extent of robustness of the MDPDE increases with increasing values of $\alpha>0$.
	As a consequence, the same is also expected for the proposed DPD-SIS (or DPD-ISIS) with 
	increasing $\alpha>0$.
	
	\begin{remark}
		Although we have seen that the robustness increases as $\alpha>0$ increases, 
		we cannot use the larger values of $\alpha$ in every cases. 
		This is because, when there is no outlier (pure data) with respect to the assumed (marginal) regression model, 
		the asymptotic variance of the MDPDE   $\widehat{\beta}_j^{M\alpha}$ is known to increase with increasing $\alpha$ values
		and hence their asymptotic relative efficiency (ARE) compared to the (most efficient but non-robust) OLS/MLE
		decreases as $\alpha$ increases (see Figure \ref{FIG:ARE}). 
		Therefore, in summary, the tuning parameter $\alpha$ provides a trade-off between efficiency of the MDPDE under pure data 
		and its robustness under data contamination (which is the case with most common robust inference approaches)
		so that we need to choose $\alpha$ carefully for any practical application, ideally depending on 
		the amount of expected contamination in the data.  
		In the following section (Section \ref{SEC:simulation}), we will empirically investigate the performance of the proposed 
		MDPDE-based variables screening procedures, DPD-SIS and DPD-ISIS, under various situations
		and accordingly provide more insights and suggestions on the choice of $\alpha$ in DPD-SIS (or DPD-ISIS)
		in the subsequent Section \ref{SEC:choice}.
	\end{remark}

	\section{Simulation Studies}
	\label{SEC:simulation}
	
	\subsection{Experimental Plans}
	\label{SEC:sim_setup}
	
	We have performed extensive simulation studies to study and illustrate the performance of our proposed DPD based screening procedures.
	For each set-up we have simulated a random sample of size $n$ from a linear regression model (\ref{EQ:LRM}) of dimension $p \gg n$
	where the $(p-1)$ covariates, except the intercept, are generated from a multivariate normal distribution having mean vector 
	$\boldsymbol{0}$ and some specified covariance matrix, say $\Sigma_x$. 
	After generating covariate values and error components for some fixed $\sigma^2$, 
	the responses are computed based on specified true values of the regression coefficient $\boldsymbol{\beta}\in \mathbb{R}^p$;
	these true values are taken to be sparse with only the first  $s=5$ components being non-zero 
	and the rest being zero. So, other than the intercept, only four covariates are significantly related with 
	the response variable and the rest are noise covariates in all our simulation set-ups.
	Additionally, to study the robustness, a part (say, $100\epsilon\%$) of the samples are contaminated
	through an appropriate contamination scheme. 
	All the parameters in the simulations are considered as follows.
	\begin{itemize}
		\item Two possible sample sizes are considered; $n=50$ and $n=100$.
		For each case, the model dimension is taken as $p=5000$ to mimic the common ultra-high dimensional set-ups appearing in real life. 	Recall $s=5$.
		These set-ups are clearly more extreme with regard to dimensionality 
		compared to the set-ups studied 
		in the SIS literature, but we believe they are closer to the true scenarios in practical Omics data analysis.
		
		\item The first $s=5$ non-zero coefficient values are all taken as 1. Three different values of the error variance
		are considered; given by $\sigma=0.2, 1,2$, which yield three types of signal strengths 
		having different signal-to-noise (SN) values. We refer to them, respectively, as \textit{strong} (SN=5), 
		\textit{moderate} (SN=1) and \textit{weak} (SN=0.5) signals.
		
		\item Different correlation structures are considered among the covariates via different $\Sigma_x$. In particular, 
		we consider \textit{independent}  covariates with $\Sigma_x$ being identity,
		and two types of correlated cases with the $(i,j)$-th elements of $\Sigma_x$ being $\rho^{|i-j|}$, and  $\rho I(i\neq j)$. 
		We will refer to these two cases, respectively, as the case of 	\textit{partially correlated} and 
		\textit{strongly correlated} covariates. 
		Several values of $\rho$ have been studied but the SIS performance results corresponding to only $\rho=0.5$ (in both cases)
		are reported in the paper for brevity.
		
		\item We have also studied different types of contamination schemes which all yield similar (in spirit) results.
		Hence, for brevity, we present the results for one particular contamination scheme 
		where the responses are contaminated by replacing its value $y$ (say) by $(y-30)$.
		The contamination proportion is taken as $\epsilon= 0.05, 0.1, 0.2$, 
		resulting in mild, moderate and heavy contaminations, respectively.
	\end{itemize}
	
	For each simulation set-up, we have applied the proposed DPD-SIS procedure to select the important variables
	and different performance measures are computed in order to study the results.
	The whole process is replicated 300 times to report some stable summary of the performance measures.
	In particular, the performance measures considered are
	\begin{eqnarray}
	\begin{array}{ll}
	\mbox{IC} & : \mbox{  Indicator if all (4) important covariates are selected in a model of size ($n-1$)}.\\
	\mbox{TP} & : \mbox{ The number of true positives selected when a model of size ($n-1$) is chosen.}\\
	\mbox{MMS} & : \mbox{ Minimum model size required to select all (4) important covariates.}
	\end{array}\nonumber
	\end{eqnarray}
	Note that the average IC over all 300 replications yields 
	the percentage of times the full model is selected in a model of size ($n-1$). This is reported in the tables.
	However, for a deeper understanding,
	resulting values of TP and MMS from 300 replications are presented in terms of box-plots and histograms, respectively.
	Additionally, a run-time comparison is provided towards the end of this section.
	
	Along with studying our proposed DPD-SIS, the above performance measures are also used to compare our proposal 
	with existing parametric and nonparametric competitive screening procedures.
	In this paper, we have considered the following competitive methods,
	where the first two are usual SIS approaches (non-robust) and 
	the remaining four are robust non-parametric extensions of the SIS available in the literature.
	\begin{itemize}
		\item \textbf{SIS}: The usual SIS of \citet{Fan/Lv:2008} which use the Pearson correlation between the response and covariates
		for screening.
		
		\item \textbf{Reg-SIS}: The version of the usual SIS obtained from a marginal regression model, as in \cite{Fan/Song:2010},
		but considering the error variance as a parameter in the model and estimating it (along with the regression coefficients)
		using the maximum likelihood approach. This is indeed the same as DPD-SIS($\alpha=0$).
		
		\item \textbf{Rank-SIS}: A robust extension of SIS obtained by using non-parametric rank correlation 
		in place of Pearson correlation in SIS \citep{Li/etc:2012a}.
		
		\item \textbf{GK-SIS}: A robust extension of SIS obtained by using a robust correlation measure proposed 
		by \citet{Gnanadesikan/Kettenring:1972} in place of Pearson correlation \citep{Gather/Guddat:2008}.
		
		\item \textbf{dCor-SIS}: A robust extension of SIS obtained by using a distance based correlation measure
		from \citet{Szekely/etc:2007} in place of Pearson correlation \citep{Wang/etc:2017,Li/etc:2012b}.
		
		\item \textbf{MCP-SIS}: A robust extension of SIS based on a robust measure of association, 
		namely the median of component-wise products (MCPs) introduced by \citet{Mu/Xiong:2014},
		which is used to rank the covariate importance.
	\end{itemize}
	Another non-parametric robust screening procedure is available in the literature,
	based on the bivariate winsorized (BW) correlation estimator of \citet{Khan/etc:2007} 
	in place of the usual correlation in SIS; we have not considered this BW based SIS, 
	since \citet{Mu/Xiong:2014} have already shown it to have similar performance as the MCP-SIS considered here.
	
	In the following two subsections, we give the results from our simulation studies
	with pure and contaminated data, respectively, for studying the performance of different SIS approaches.
	Another set of simulation results illustrating the performance of DPD-ISIS will be discussed 
	later in Subsection \ref{SEC:sim_ISIS}.

	\subsection{Performance of the DPD-SIS without contamination}
	\label{SEC:sim_pure}
	
	Let us first illustrate the performance of the DPD-SIS under pure data containing no outliers.
	For all the simulation set-ups without contamination as described in the previous subsection, 
	the percentage of times the full (correct) model is selected (average IC) by different SIS approaches with target model size $d=n-1$
	is reported in Table \ref{TAB:pure_PIC}. One can immediately see that, as expected, 
	all the SIS methods fails in case of strongly correlated covariates (except for large sample sizes and strong signal strength)
	and we need to use appropriate ISIS is such cases; 
	we will illustrate the performance of our proposed DPD-ISIS later in Section \ref{SEC:sim_ISIS}.
	For the other two types of covariates, 
	the performance of our proposed DPD-SIS under pure data deteriorate slightly with increasing values of $\alpha$
	(due to the loss in efficiency of MDPDE under pure data), but the DPD-SIS at $\alpha=0.1, 0.3$ are  pretty much comparable with 
	the usual SIS is most cases and also significantly better compared to the existing non-parametric robust SIS approaches.
	For all the partially correlated cases as well as independent cases with moderate to strong signals and $n=100$,
	the DPD-SIS  provides the correct full model in over 90\% of the replications which decreases as the signal strength becomes weaker or sample size becomes smaller.
	Among the two types of covariates, the performance is far better when some amount of correlation is present compared to 
	the fully independent covariates when we have weaker signal strength and/or smaller sample sizes.
	The good behavior of the methods in the case with partially correlated data is caused by the way we simulated our data, 
	with a cluster of important variables at the start of the $X$-sequence. 
	When these important variables are randomly distributed in the sequence, the results are less good (as expected). 
	This holds for all methods, but their relative behavior is again observed to be the same as in the present case. 
	So, to keep our focus on comparison between the models, 
	we have not presented the results for randomly distributed important covariates for brevity. 

\begin{figure}[!h]
		\centering
		\subfloat[Strong Signal; $n=50$]{
			\includegraphics[width=0.33\textwidth]{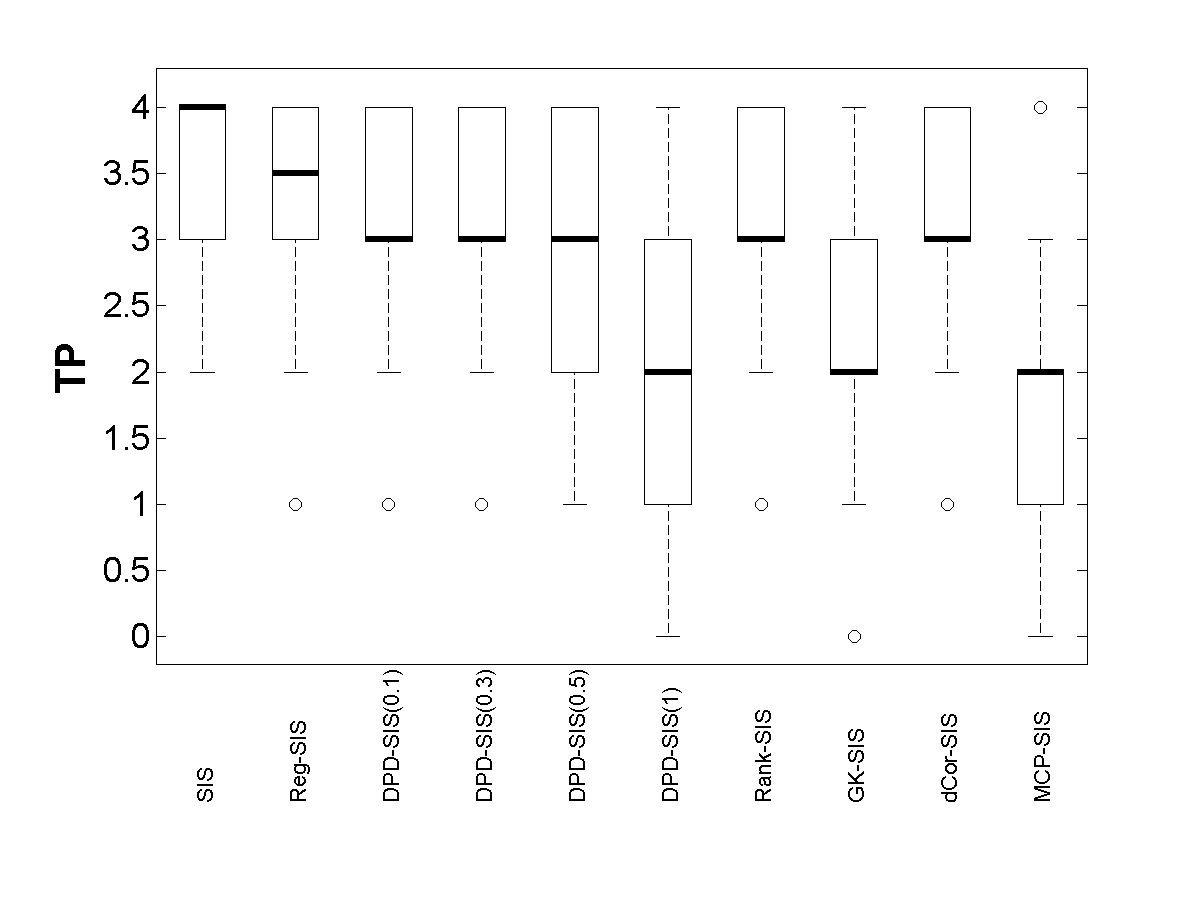}
			\label{FIG:TP_Set1_Ss_n50_cont00}}
		\subfloat[Moderate Signal; $n=50$]{
			\includegraphics[width=0.33\textwidth]{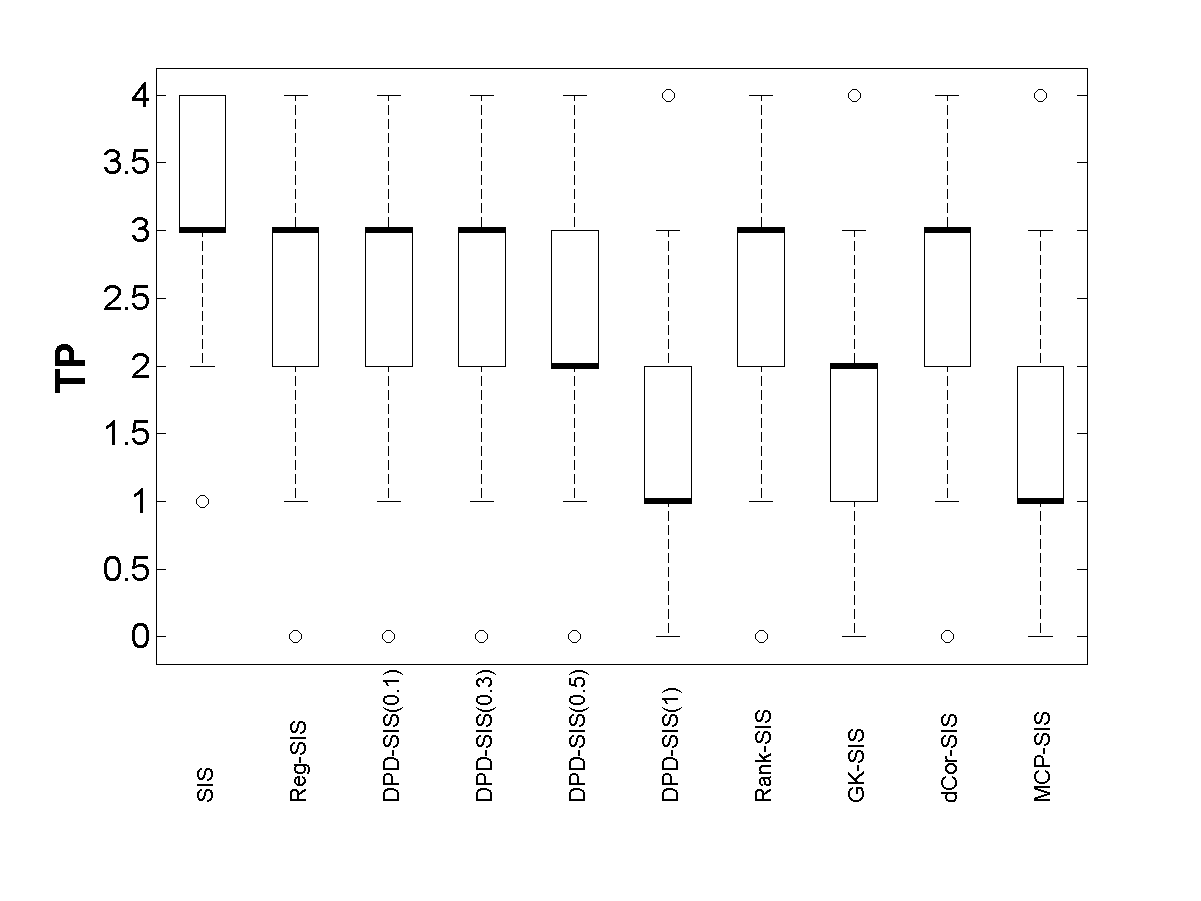}
			\label{FIG:TP_Set1_Ms_n50_cont00}}
		\subfloat[Weak Signal; $n=50$]{
			\includegraphics[width=0.33\textwidth]{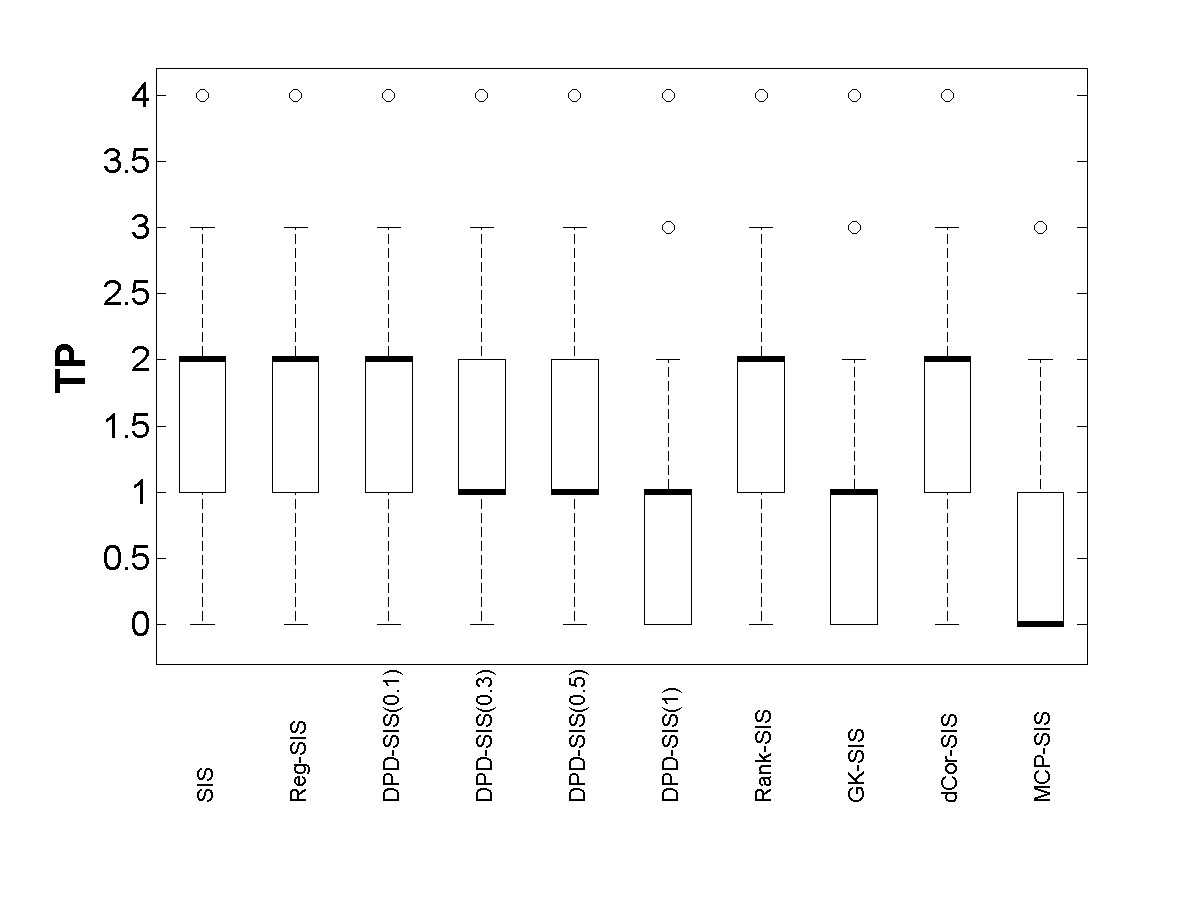}
			\label{FIG:TP_Set1_Ws_n50_cont00}}
		\\	
		\subfloat[Strong Signal; $n=100$]{
			\includegraphics[width=0.33\textwidth]{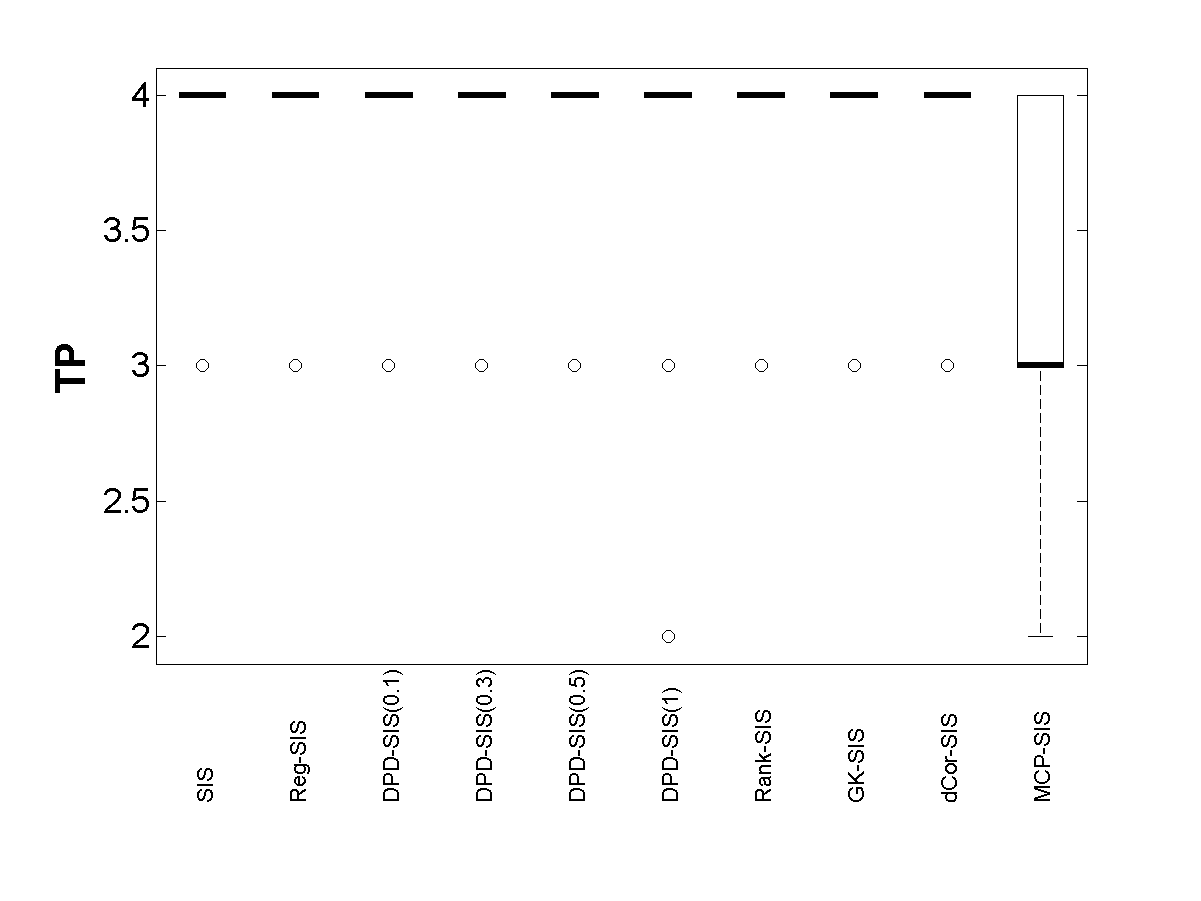}
			\label{FIG:TP_Set1_Ss_n100_cont00}}
		\subfloat[Moderate Signal; $n=100$]{
			\includegraphics[width=0.33\textwidth]{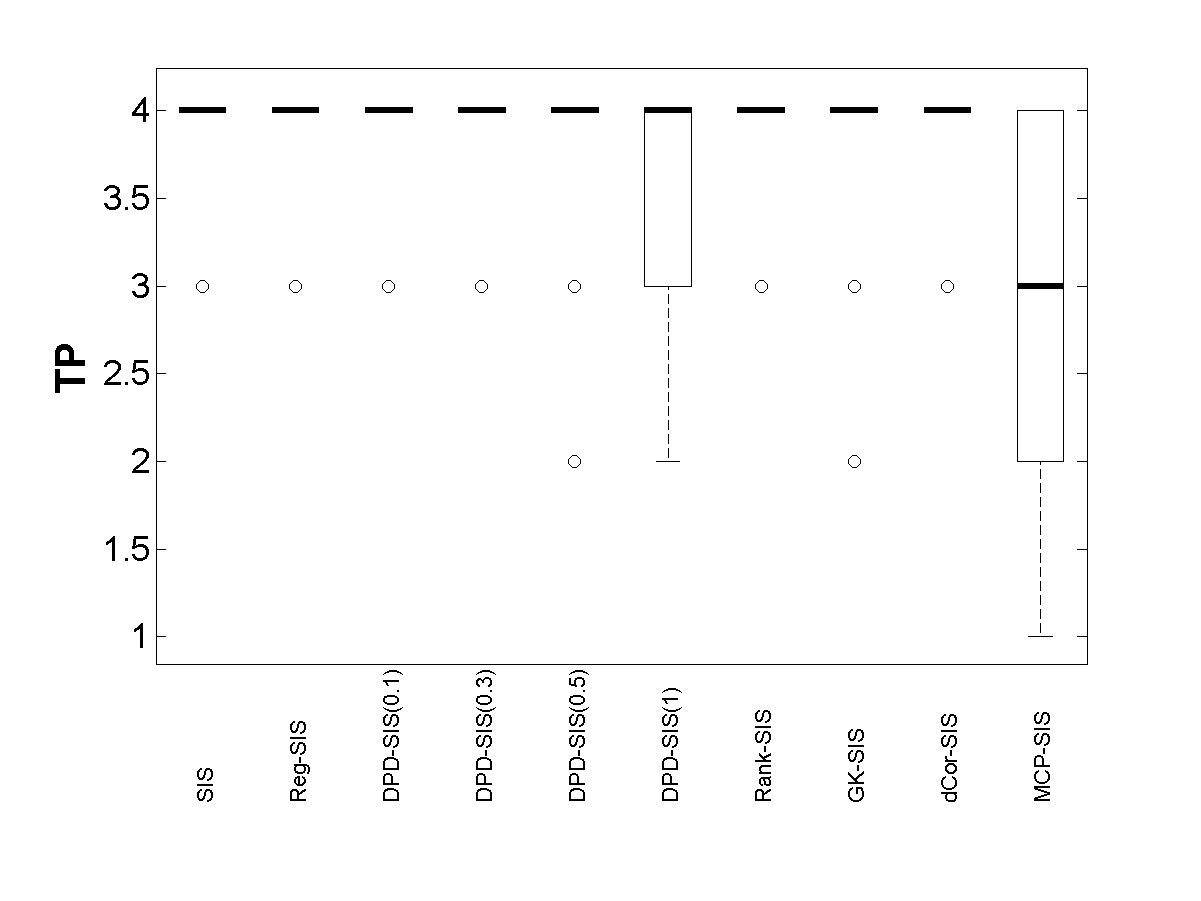}
			\label{FIG:TP_Set1_Ms_n100_cont00}}
		\subfloat[Weak Signal; $n=100$]{
			\includegraphics[width=0.33\textwidth]{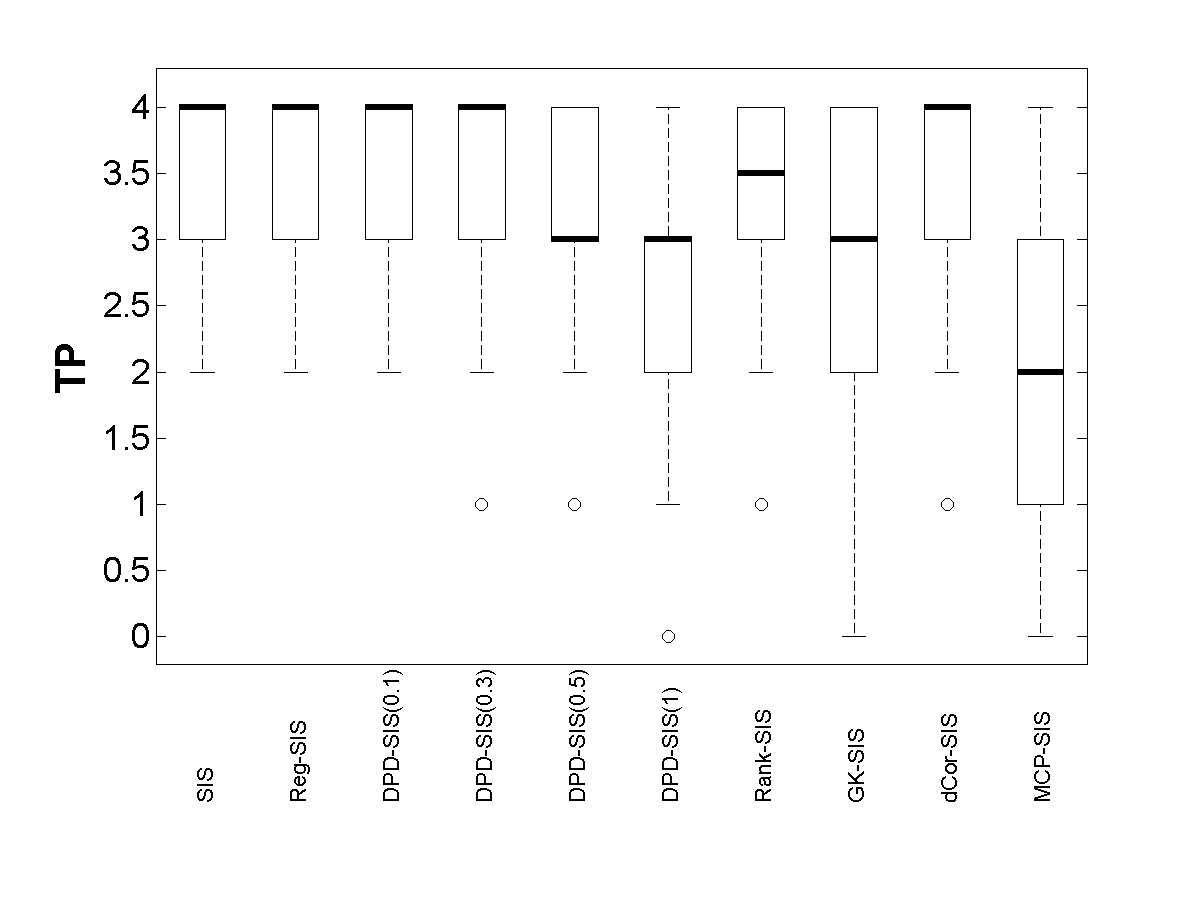}
			\label{FIG:TP_Set1_Ws_n100_cont00}}
		
		\caption{Box-Plots of the true-positives (TP) obtained by different SIS approaches with target model size $d=n-1$
			for independent covariates with pure data}
		\label{FIG:TP_pure_indep}
	\end{figure}

	
	The methods can be compared further via the other performance measures, TP and MMS.
	With regard to TP, our simulations show that the median of the true positives selected by the usual SIS, Reg-SIS  and our DPD-SIS at $\alpha\leq 0.5$ 
	with a target model size of $d=n-1$ are all equal four (the true active set size) for the cases of partially correlated covariates under pure data
	and hence, they are extremely comparative in these cases (data not shown). For the independent covariate cases, the box-plots of the obtained true-positives are 
	presented in Figure \ref{FIG:TP_pure_indep} where we can see that the results are again very similar for $n=100$.
	For smaller sample size $n=50$, however, the results are not that good; the usual SIS has median true-positive values of 4, 3 and 2, respectively, 
	for  strong, moderate and weak signals, whereas the median true positives obtained by DPD-SIS at $\alpha=0.1$ are 3, 3, and 2, respectively.
	The values of true-positives generally seem to decrease with increasing $\alpha$ in DPD-SIS under pure data scenarios
	but $\alpha=0.3$ also gives very competitive results in most cases. 
	As for the other (non-parametric) robust methods, the Rank-SIS and the dCor-SIS also perform reasonably well with regard to this measure.
	
	\begin{table}[h]
	\caption{Percentage of times the full (correct) model is selected (average IC) by different SIS approaches with target model size $d=n-1$
		for pure data}%
	\centering
	\resizebox{\textwidth}{!}{
		\begin{tabular}[c]{|l|l|rr|rrrr|rrrr|}\hline
			Signal	&	Sample	&	\multicolumn{2}{c|}{Non-robust SIS}	&	\multicolumn{4}{c|}{Proposed DPD-SIS($\alpha$)}	
			&	\multicolumn{4}{c|}{Existing Robust (Non-parametric) SIS}	\\
			Strength & size ($n$) &	SIS	&	Reg-SIS	&	$\alpha=0.1$	&	$\alpha=0.3$ &	$\alpha=0.5$	&	$\alpha=1$	
			&	Rank-SIS	&	GK-SIS	&	dCor-SIS	&	MCP-SIS	\\\hline
			\hline
			\multicolumn{12}{|l|}{\underline{Independent Covariates}}\\
			Strong	&	50	&	55.0	&	50.0	&	48.3	&	41.7	&	27.7	&	8.0	&	41.7	&	12.7	&	43.7	&	1.7	\\
			&	100	&	99.0	&	99.7	&	99.7	&	98.7	&	98.3	&	90.7	&	97.7	&	93.7	&	97.7	&	49.3	\\
			Moderate	&	50	&	25.3	&	20.7	&	18.7	&	15.7	&	10.0	&	1.0	&	14.7	&	5.3	&	15.3	&	0.3	\\
			&	100	&	94.3	&	95.7	&	94.7	&	93.3	&	91.3	&	72.0	&	91.0	&	77.3	&	91.3	&	26.3	\\
			Weak		&	50	&	2.0	&	1.3	&	1.0	&	1.0	&	0.7	&	0.3	&	1.3	&	0.7	&	0.7	&	0.0	\\
			&	100	&	58.7	&	58.0	&	57.3	&	53.7	&	44.7	&	23.3	&	50.0	&	26.7	&	50.7	&	5.3	\\
			\hline
			\multicolumn{12}{|l|}{\underline{Partially Correlated Covariates with $\rho=0.5$}}\\
			Strong 	&	50	&	99.3	&	99.7	&	99.7	&	99.7	&	99.0	&	86.3	&	98.3	&	78.0	&	98.7	&	49.3	\\
			&	100	&	100.0	&	100.0	&	100.0	&	100.0	&	100.0	&	100.0	&	100.0	&	100.0	&	100.0	&	97.7	\\
			Moderate	&	50	&	98.7	&	99.3	&	99.0	&	98.7	&	97.3	&	73.7	&	95.7	&	68.0	&	97.7	&	37.7	\\
			&	100	&	99.7	&	99.7	&	99.7	&	99.7	&	99.7	&	99.7	&	99.7	&	99.7	&	99.7	&	95.4	\\
			Weak	&	50	&	86.7	&	86.0	&	85.3	&	82.3	&	75.7	&	38.3	&	79.0	&	42.0	&	80.7	&	20.3	\\
			&	100	&	99.7	&	99.7	&	100.0	&	100.0	&	100.0	&	99.7	&	99.7	&	97.3	&	100.0	&	85.0	\\
			\hline
			\multicolumn{12}{|l|}{\underline{Strongly Correlated Covariates  with $\rho=0.5$}}\\
			Strong	&	50	&	14.7	&	1.7	&	1.7	&	1.0	&	0.7	&	0.0	&	5.7	&	0.0	&	9.0	&	0.0	\\
			&	100	&	82.3	&	49.0	&	48.0	&	39.0	&	25.7	&	7.3	&	59.7	&	2.7	&	66.3	&	1.0	\\
			Moderate	&	50	&	5.0	&	1.0	&	1.0	&	0.7	&	0.7	&	0.0	&	2.3	&	0.0	&	2.7	&	0.0	\\
			&	100	&	31.7	&	24.4	&	17.4	&	4.4	&	64.6	&	64.6	&	43.0	&	1.0	&	50.0	&	0.4	\\
			Weak	&	50	&	0.7	&	0.0	&	0.0	&	0.0	&	0.0	&	0.0	&	0.0	&	0.0	&	0.3	&	0.0	\\
			&	100	&	25.0	&	10.3	&	8.7	&	8.3	&	5.3	&	1.3	&	14.0	&	0.7	&	16.0	&	0.3	\\
			\hline
	\end{tabular}}
	\label{TAB:pure_PIC}%
\end{table}

	\begin{figure}
		\centering
		\subfloat[Strong Signal; $n=100$]{
			\includegraphics[width=.97\textwidth]{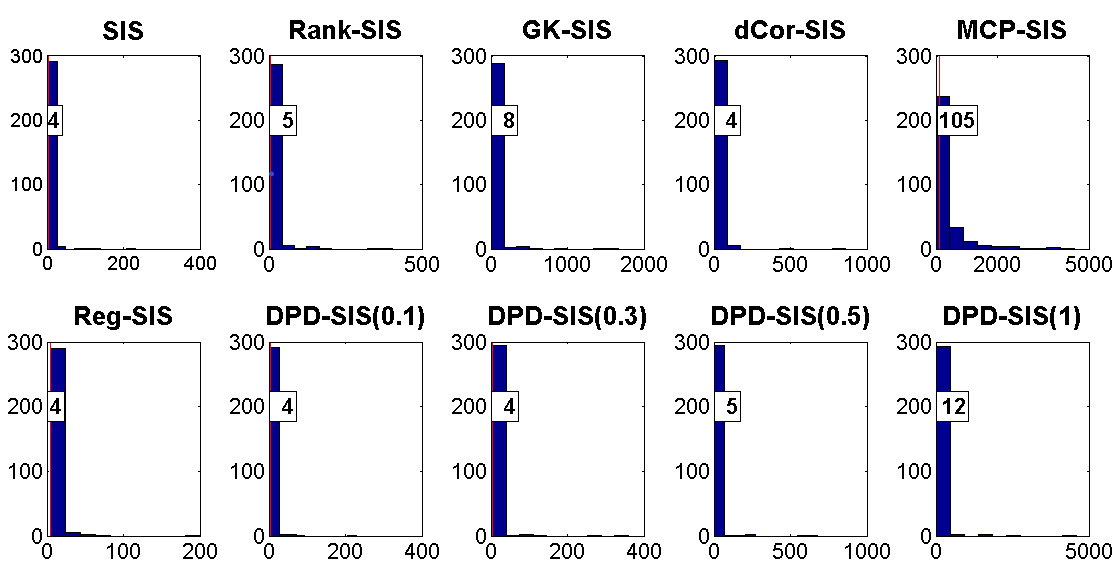}
			\label{FIG:TP_Set1_Ss_n50_cont00}}
		\\	
		\subfloat[Weak Signal; $n=50$]{
			\includegraphics[width=.97\textwidth]{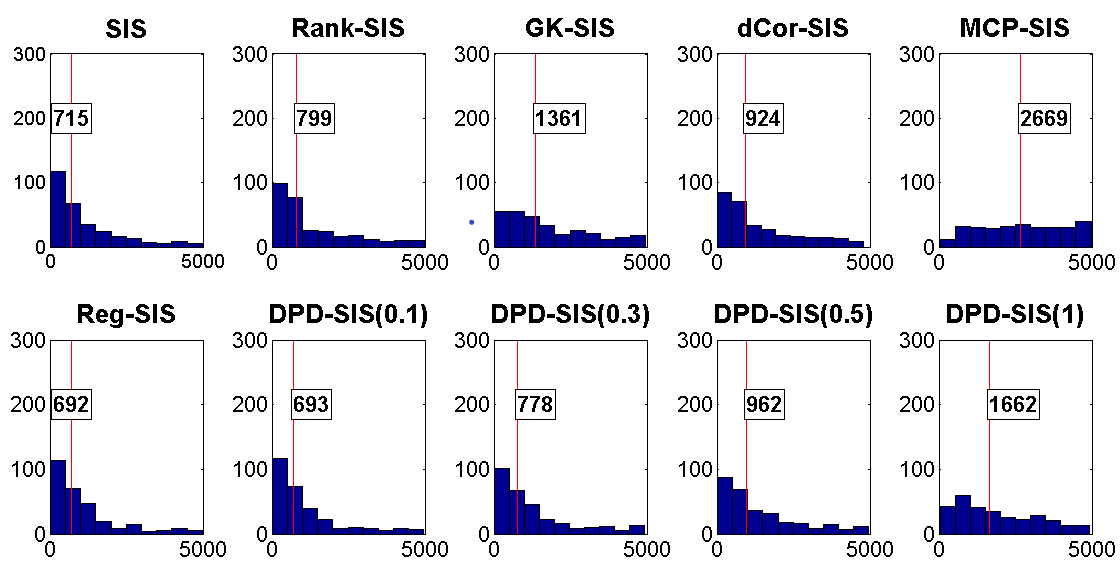}
			\label{FIG:TP_Set1_Ws_n50_cont00}}
		\caption{Histograms for the minimum target model size required by different SIS approaches to select all the four true-positives (TP)
			for independent covariate cases under pure data. The median value is also reported in each plot and marked by a red vertical line.}
		\label{FIG:MMS_pure_indep}
	\end{figure}
	
	We have further investigated MMS, the minimum target model size $(d)$ required to select all the four true positives by different SIS approaches. 
	Whenever SIS performs well, e.g., partially correlated covariates and/or strong signals,
	the MMS values are pretty low, often less than 10 with a median of about 4-6. For brevity, 
	we only present the results (histogram) on MMS for two extreme cases with independent covariates in Figure \ref{FIG:MMS_pure_indep},
	namely for strong signal with $n=100$ (one of the best performing cases) and weak signal with $n=50$ (one of the worst performing cases).
	The range (and median) of MMS differ widely in both cases but the general trend is the same 
	(which is also the same in all other cases not reported here). The the median MMS for DPD-SIS increases with increasing values of $\alpha$ 
	and are generally higher than the usual SIS in pure data; but those obtained by DPD-SIS at $\alpha=0.1, 0.3$ are very close to the values obtained 
	by the usual SIS and often significantly better than the other existing non-parametric SIS approaches.

	In summary, under pure data, usual SIS performs the best as expected, but 
	there is only a slight loss in performance by the proposed DPD-SIS with smaller values of $\alpha>0$.
	We will see next that, with this small price in case of pure data, 
	we gain significant improvement over the usual SIS by using DPD-SIS under data contamination.
	Having a parametric nature, the proposed DPD-SIS naturally  performs better than the existing non-parametric SIS approaches.

	\subsection{Performance of the DPD-SIS under data contamination}
	\label{SEC:sim_cont}

\begin{figure}[!h]
		\centering
		\subfloat[Set 1; 5\% contamination]{
			\includegraphics[width=0.33\textwidth]{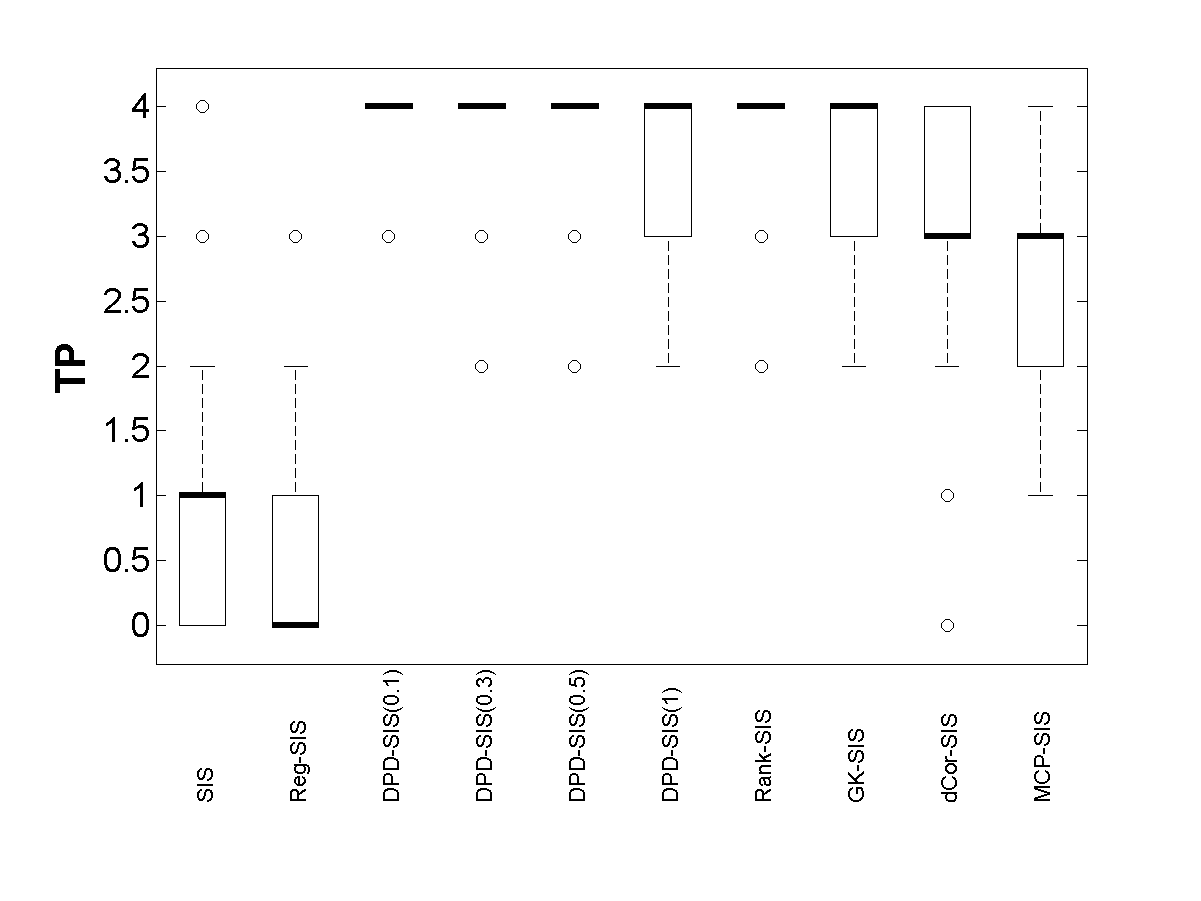}
			\label{FIG:TP_Set1_Ss_n50_cont00}}
		\subfloat[Set 1; 10\% contamination]{
			\includegraphics[width=0.33\textwidth]{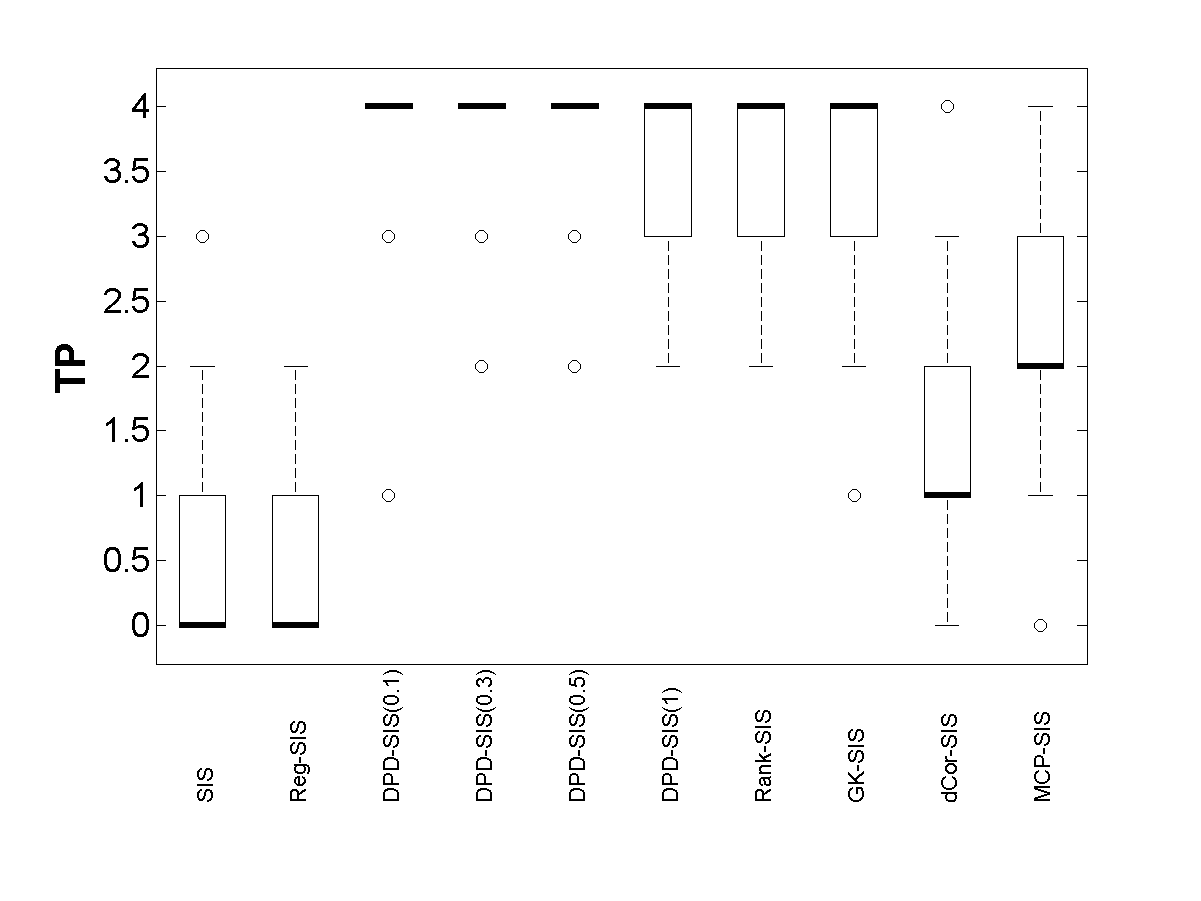}
			\label{FIG:TP_Set1_Ms_n50_cont00}}
		\subfloat[Set 1; 20\% contamination]{
			\includegraphics[width=0.33\textwidth]{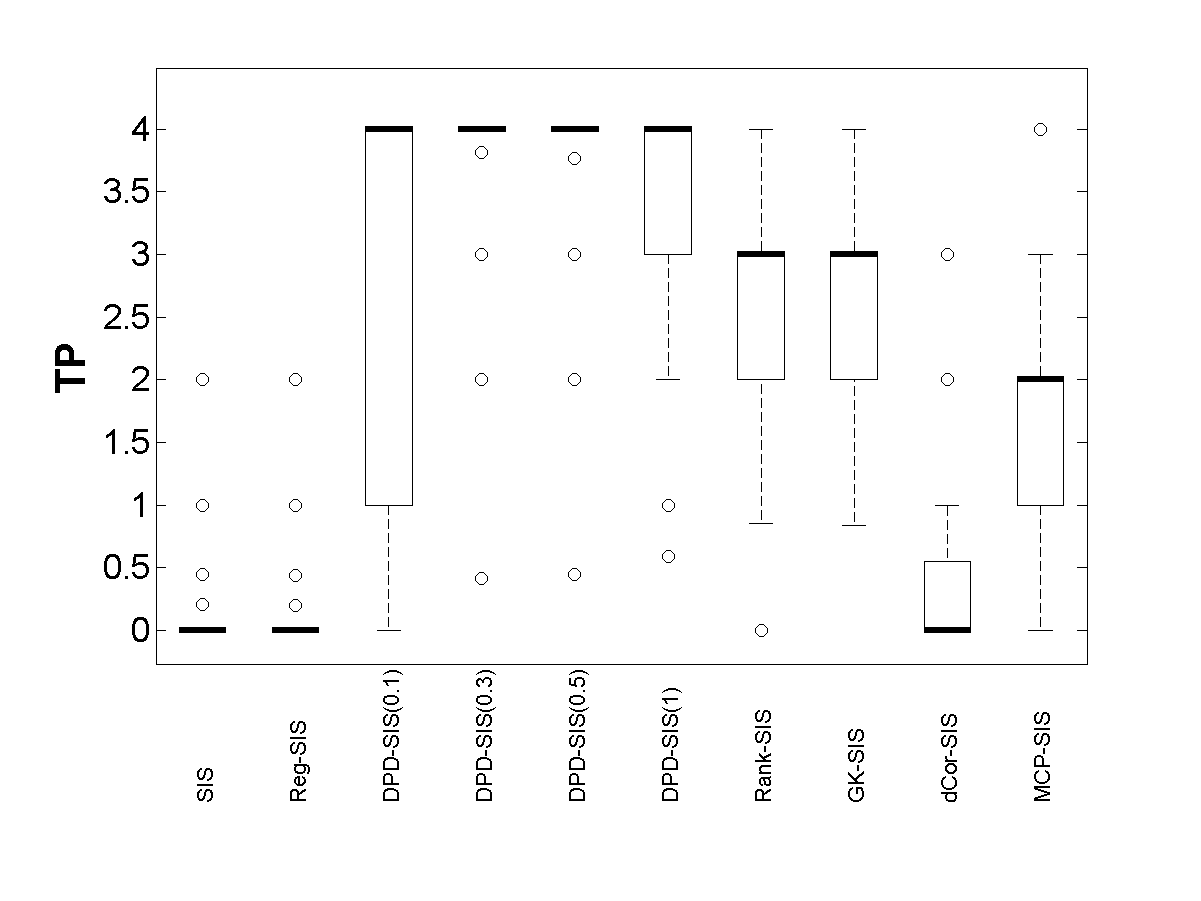}
			\label{FIG:TP_Set1_Ws_n50_cont00}}
		\\	
		\subfloat[Set 2; 5\% contamination]{
			\includegraphics[width=0.33\textwidth]{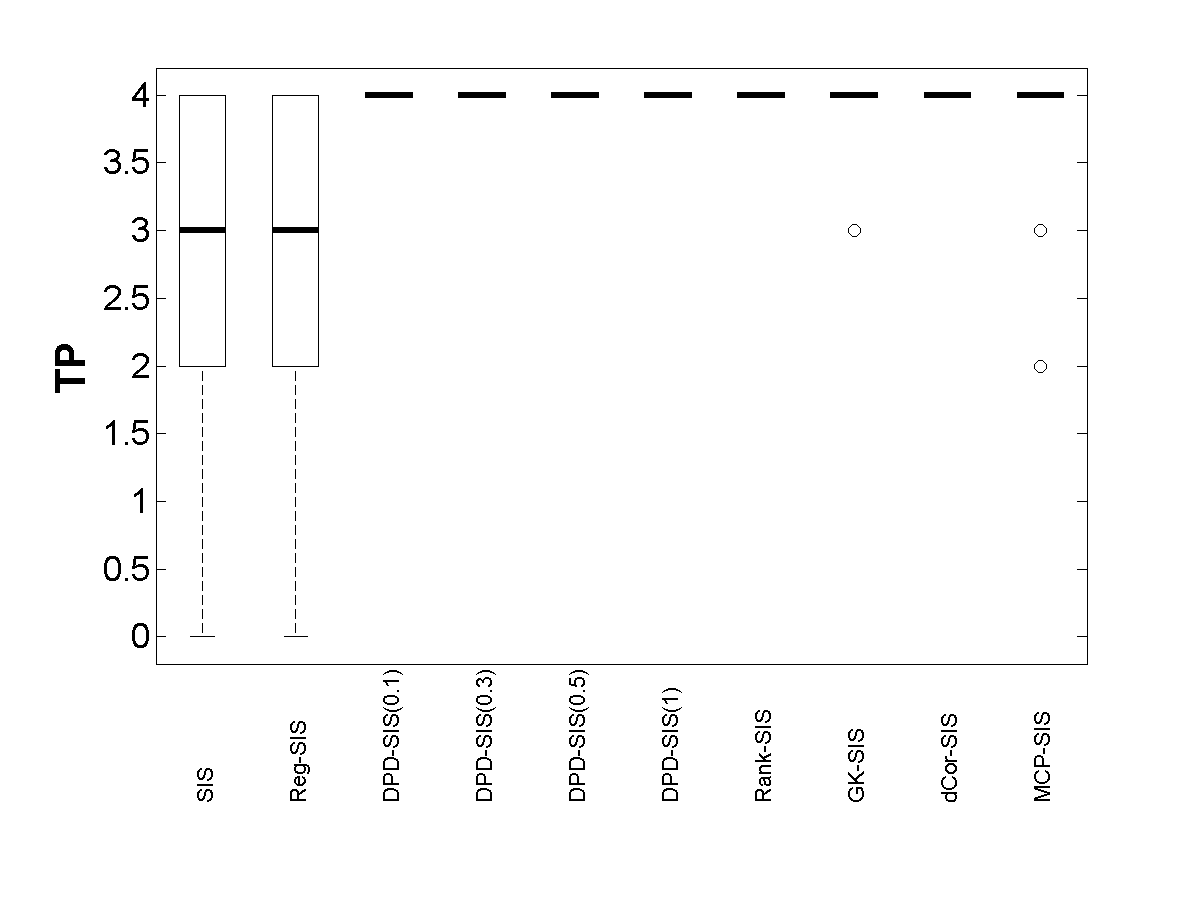}
			\label{FIG:TP_Set1_Ss_n50_cont00}}
		\subfloat[Set 2; 10\% contamination]{
			\includegraphics[width=0.33\textwidth]{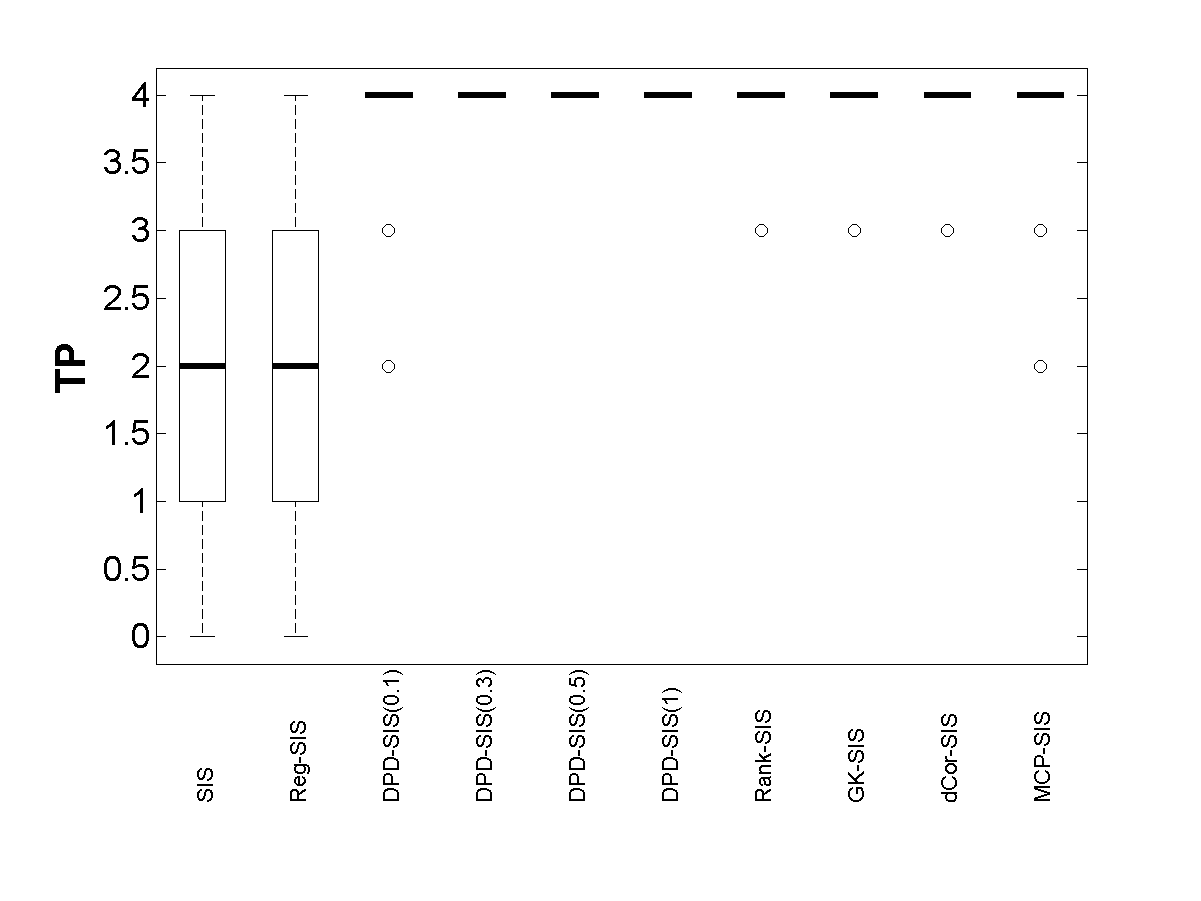}
			\label{FIG:TP_Set1_Ms_n50_cont00}}
		\subfloat[Set 2; 20\% contamination]{
			\includegraphics[width=0.33\textwidth]{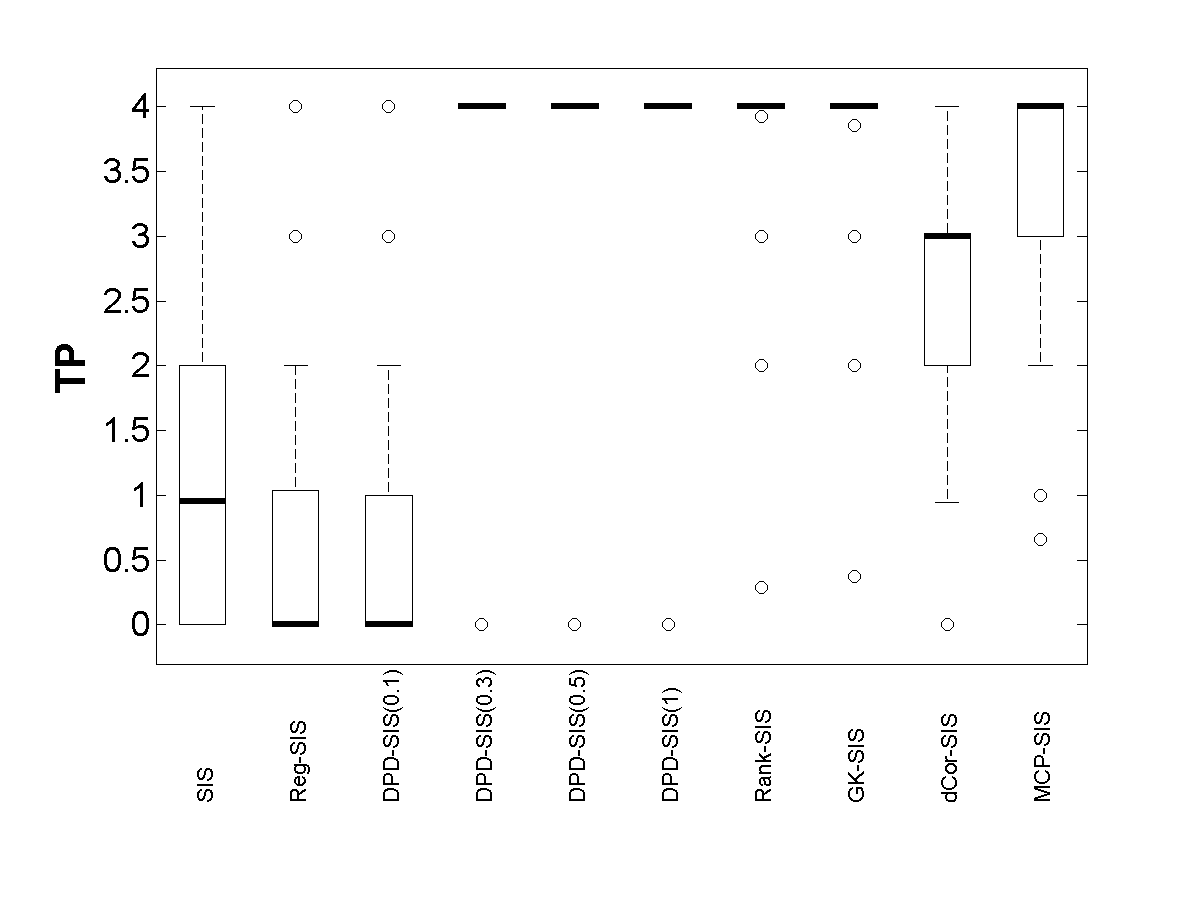}
			\label{FIG:TP_Set1_Ws_n50_cont00}}
		\caption{Box-Plots of the true-positives (TP) obtained by different SIS approaches with target model size $d=n-1$
			for independent covariates (Set 1) and partially correlated covariates (Set 2) with $n=100$ and moderate signal strength under different amount of contamination in data.}
		\label{FIG:TP_cont_n100}
	\end{figure}

	Let us now illustrate the performance of our DPD-SIS under data contamination
	and investigate the claimed improvements over the existing SIS and non-parametric robust SIS approaches.
	Due to the similarity in the patterns of results across all the cases considered 
	(the only difference being in the magnitude of the performance measures, as in the pure data cases),
	we here only present the  results for a representative case of $n=100$ and moderate signal strength for both 
	the independent and partially correlated covariates. For these representative cases, 
	the percentage  of times the full model is selected and the box-plots of the actual numbers of true-positives 
	selected by different SIS approaches with target model size $d=n-1$ are presented in Table \ref{TAB:PIC_cont}
	and Figure \ref{FIG:TP_cont_n100}, respectively.

	\begin{table}[h]
		\caption{Percentage of times the full (correct) model is selected (average IC) by different SIS approaches 
			with target model size $d=n-1$ for contaminated data with sample size $n=100$, moderate signal strength
			and different contamination proportion ($\epsilon$). Corresponding values for pure data are also given for comparison.}%
		\centering
		\resizebox{\textwidth}{!}{
			\begin{tabular}[c]{|l|rr|rrrr|rrrr|}\hline
				&	\multicolumn{2}{c|}{Non-robust SIS}	&	\multicolumn{4}{c|}{Proposed DPD-SIS($\alpha$)}	
				&	\multicolumn{4}{c|}{Existing Robust (Non-parametric) SIS}	\\
				$100\epsilon\%$ &	SIS	&	Reg-SIS	&	$\alpha=0.1$	&	$\alpha=0.3$ &	$\alpha=0.5$	&	$\alpha=1$	
				&	Rank-SIS	&	GK-SIS	&	dCor-SIS	&	MCP-SIS	\\\hline
				\hline
				\multicolumn{11}{|l|}{\underline{Independent Covariates}}\\
				0\% 	&	94.3	&	95.7	&	94.7	&	93.3	&	91.3	&	72.0	&	91.0	&	77.3	&	91.3	&	26.3	\\
				5\%		&	0.3	&	0.0	&	94.3	&	91.0	&	89.0	&	73.7	&	77.3	&	63.0	&	32.7	&	18.0	\\
				10\%	&	0.0	&	0.0	&	91.3	&	89.0	&	85.7	&	70.3	&	55.3	&	51.3	&	2.7	&	12.3	\\
				20\%	&	0.0	&	0.0	&	59.3	&	82.5	&	77.5	&	64.6	&	25.1	&	23.1	&	0.0	&	3.7	\\
				\hline
				\multicolumn{11}{|l|}{\underline{Partially Correlated Covariates with $\rho=0.5$}}\\
				0\%		&	99.7	&	99.7	&	99.7	&	99.7	&	99.7	&	99.7	&	99.7	&	99.7	&	99.7	&	95.4	\\
				5\%		&	31.0	&	29.7	&	100.0	&	100.0	&	100.0	&	100.0	&	100.0	&	99.7	&	100.0	&	94.3	\\
				10\%	&	6.3	&	5.7	&	93.3	&	100.0	&	100.0	&	100.0	&	99.7	&	97.3	&	96.7	&	85.7	\\
				20\%	&	2.0	&	1.4	&	1.7	&	99.7	&	99.7	&	99.7	&	92.8	&	85.8	&	21.7	&	60.3	\\
				\hline
		\end{tabular}}
		\label{TAB:PIC_cont}%
	\end{table}

	It can be noted that the usual SIS as well as its modification, the Reg-SIS, performs extremely poorly under 
	any amount of contamination. Even at only 5\% contamination, they select all the true positives in only about 30\% of the cases with partially correlated data and almost never for the independent covariates, 
	although these numbers were 99.7\% and about 94-96\%, respectively, under no contamination. 
	As the contamination proportion increases, their performance becomes even worse
	and the same poor performance can also be seen in terms of the median number of true positives selected 
	by these methods in Figure \ref{FIG:TP_cont_n100}.
	Our proposed DPD-SIS with $\alpha>0$ shows a much more stable performance under data contamination.
	In terms of percentages of full model selection, DPD-SIS with $\alpha\approx 0.3$ yields the best performance under heavy contamination (20\%)
	and are quite competitive to the choice of $\alpha=0.1$ also at milder contamination of 5\%.
	A similar improved performance of our DPD-SIS is observed over the usual SIS or Reg-SIS in terms of selected true positives as well.
	More interestingly, our DPD-SIS with $\alpha\in[0.3, 0.5]$ often outperforms the existing non-parametric robust SIS approaches 
	and the improvement becomes more significant at higher contamination level and for the cases of independent samples (or weaker signals).
	For the partially correlated covariates, the non-parametric Rank-SIS and GK-SIS performs quite good with a median true positive equal to four 
	(the actual active set size) but have an overall worse performance (more outlying cases with lower number of true-positives selected) 
	compared to DPD-SIS with moderate $\alpha$ values.
	
	\begin{figure}[!h]
		\centering
		\subfloat[Independent Covariates]{
			\includegraphics[width=.96\textwidth]{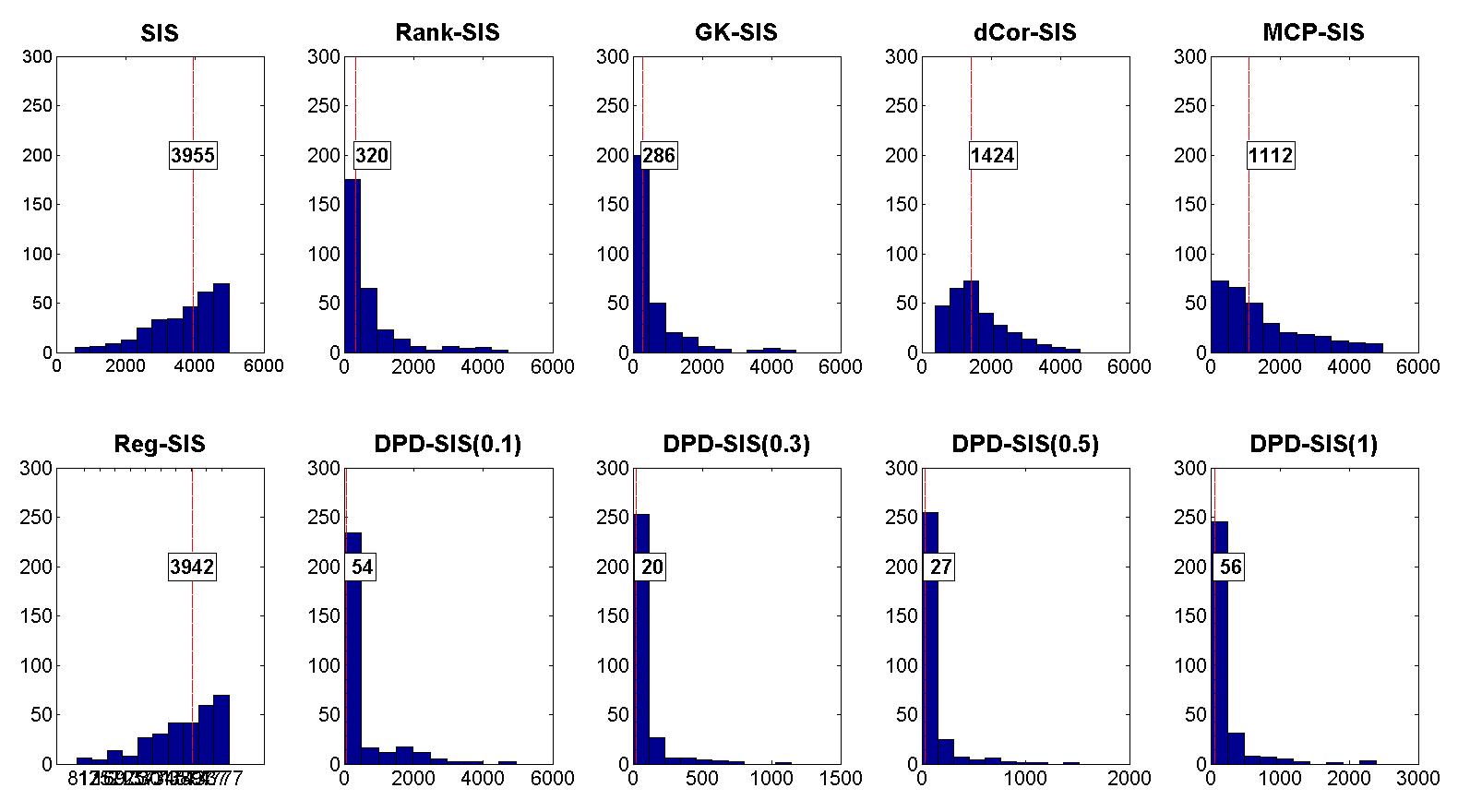}
			\label{FIG:TP_Set1_Ss_n50_cont00}}
		\\	
		\subfloat[Parially Correlated Covariates]{
			\includegraphics[width=.96\textwidth]{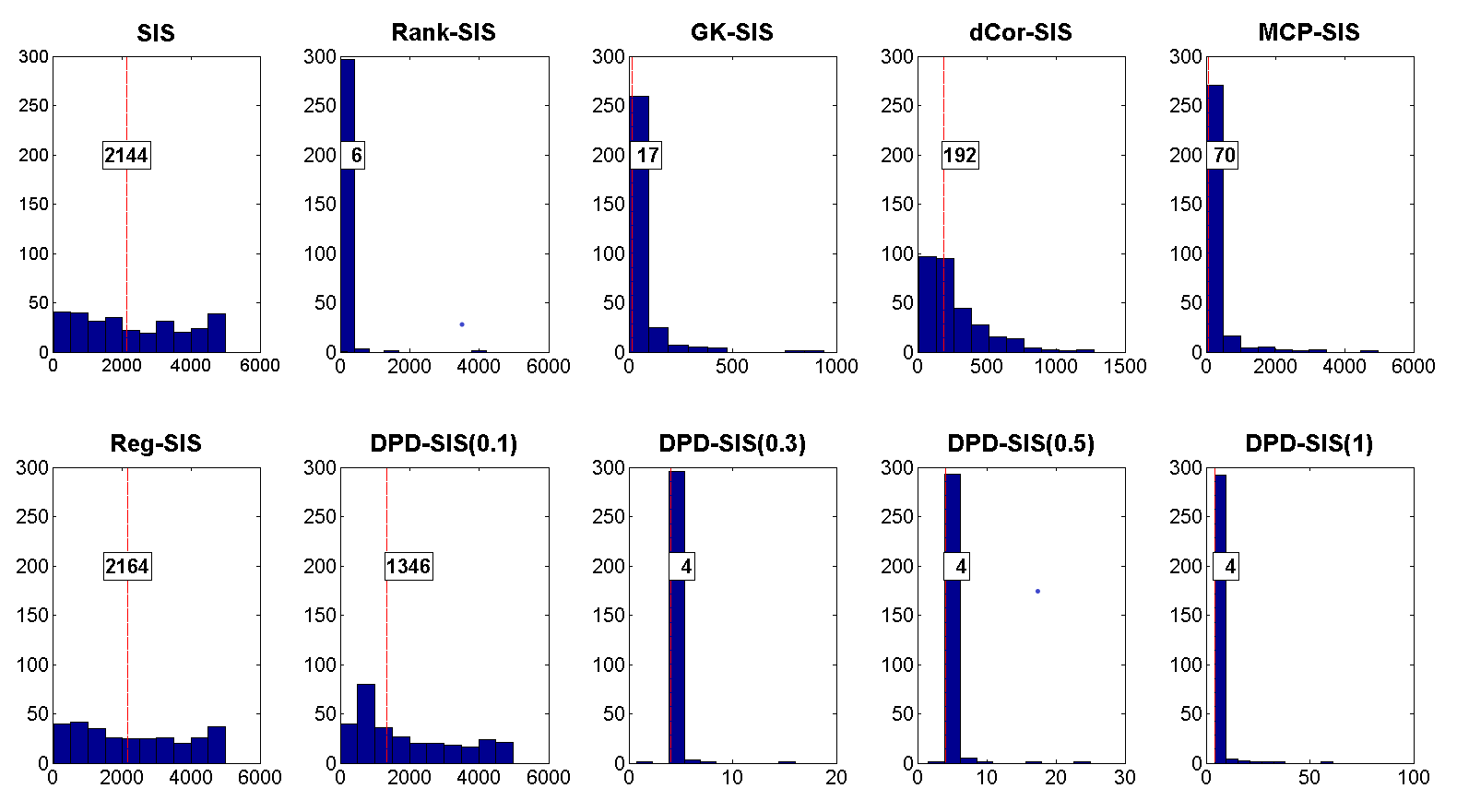}
			\label{FIG:TP_Set1_Ws_n50_cont00}}
		\caption{Histograms for the minimum target model size required by different SIS approaches to select all four true-positives
			for $n=100$ and moderate signal strength under 20\% contamination in data. 
			The median value is also reported in each plot and marked by a red vertical line.}
		\label{FIG:MMS_cont_n100}
	\end{figure}

	We have also studied the minimum target model size (MMS) required to select all four true positives under contamination
	which further illustrates the huge advantage of the proposed DPD-SIS over existing SIS approaches. 
	For brevity, the results for 20\% contamination under the representative cases are shown in Figure \ref{FIG:MMS_cont_n100}.
	Note that the median values of the MMS required by the usual SIS and Reg-SIS are of the order 3950 and 2150, respectively,
	for the independent and partially correlated covariates. These become heavily improved by the existing non-parametric robust SIS approaches
	with Rank-SIS and GK-SIS yielding better performance compared to the other two. 
	But, still for these two cases of independent or partially correlated covariates, 
	they reach the minimum median MMS of 286 (by GK-SIS) and six (by Rank-SIS), respectively.
	Our proposed DPD-SIS with $\alpha\geq 0.3$ clearly outperforms all these existing methods yielding even lower  values of MMS
	with a median of four (the minimum possible value) for the partially correlated case. 
	For the independent covariates the improvement is even more significant with the best performance of DPD-SIS at $\alpha=0.3$
	which provides a median MMS of 20 only (in comparison with the minimum value of 286 obtained by existing approaches).
	The results for all other simulation experiments, not presented here for brevity, have indicated the similar advantages of our proposed DPD-SIS
	under different types and amounts of data contamination with the improvements being larger for the more vulnerable cases of heavy contamination
	or weaker signal strength or smaller sample sizes.
	
	\begin{remark}[Runtime Comparison]
		One important advantage of SIS in the context of ultra-high dimensional data is its high computational speed. 
		Although MDPD is generally computationally intensive in large regression models, our DPD-SIS is pretty fast since it involves
		the computation of the MDPDE of three parameters only in each marginal regression model.
		Across our simulation studies, performed on a laptop having 32GB RAM and using parallel processing with 7 cores,
		the DPD-SIS for each sample with $p=5000$ variables is seen to have a median runtime of 5-6 seconds, 
		whereas the median runtime for the usual SIS (and Reg-SIS) for the same set-up is about 3 (and 5.5) seconds.
		The median runtime for the existing non-parametric SIS approaches are about 3-3.5 seconds for Rank-SIS, GK-SIS and MCP-SIS
		whereas for dCor-SIS it is approximately 5.2 seconds. So, we can clearly see that, on average, 
		our DPD-SIS takes slightly more time than the other existing SIS approaches but it is still pretty fast, and fast enough to be useful 
		for ultra-high dimensional set-ups considering its significantly improved robustness advantages as seen above.
	\end{remark}
	
\subsection{Performances of the DPD-ISIS for strongly correlated covariates}
\label{SEC:sim_ISIS}
	
Let us now study the performance of the DPD-ISIS for the case of strongly correlated covariates	
where the variance of each covariate is taken as 1 and the correlations between any pair of covariates are taken to be equal ($\rho$).
Since we have already seen significantly better performance of the DPD based (non-iterative) screening over the existing non-parametric screening approaches,
the iterative approach is only examined for DPD-ISIS described in Section \ref{SEC:DPD_ISIS}
along with the Van-ISIS \citep{Saldana/Feng:2018} as the benchmark of comparison. 
As before, we take different sampel sizes $n$, $p=5000$ and the first $s=5$ non-zero coefficients to be $\tau$
and contaminate a certain percentage of the responses by replacing its value (say $y$) by $(y-30)$.   
It is to be noted that, under such strongly correlated case, certain restrictions are required even for the usual ISIS to select all the important variables
in terms of high signal-to-noise and $n/p$ ratios for given values of $\rho$ \citep{Fan/Li:2001,Fan/Lv:2008}. 
So, we have investigated the performance of DPD-ISIS  for different values of $n$, $\rho$ and $\tau$ by fixing the error variance $\sigma^2=1$.
Considering the similarity in the pattern of the simulation results, for brevity,
we only present the results for the case $\rho=0.5$, $n=100$ and $\tau=1, 5$ (corresponding to SNR being 1 and 5, respectively).

\begin{figure}[!h]
\centering
	\subfloat[SNR=1; No contamination]{
		\includegraphics[width=0.33\textwidth]{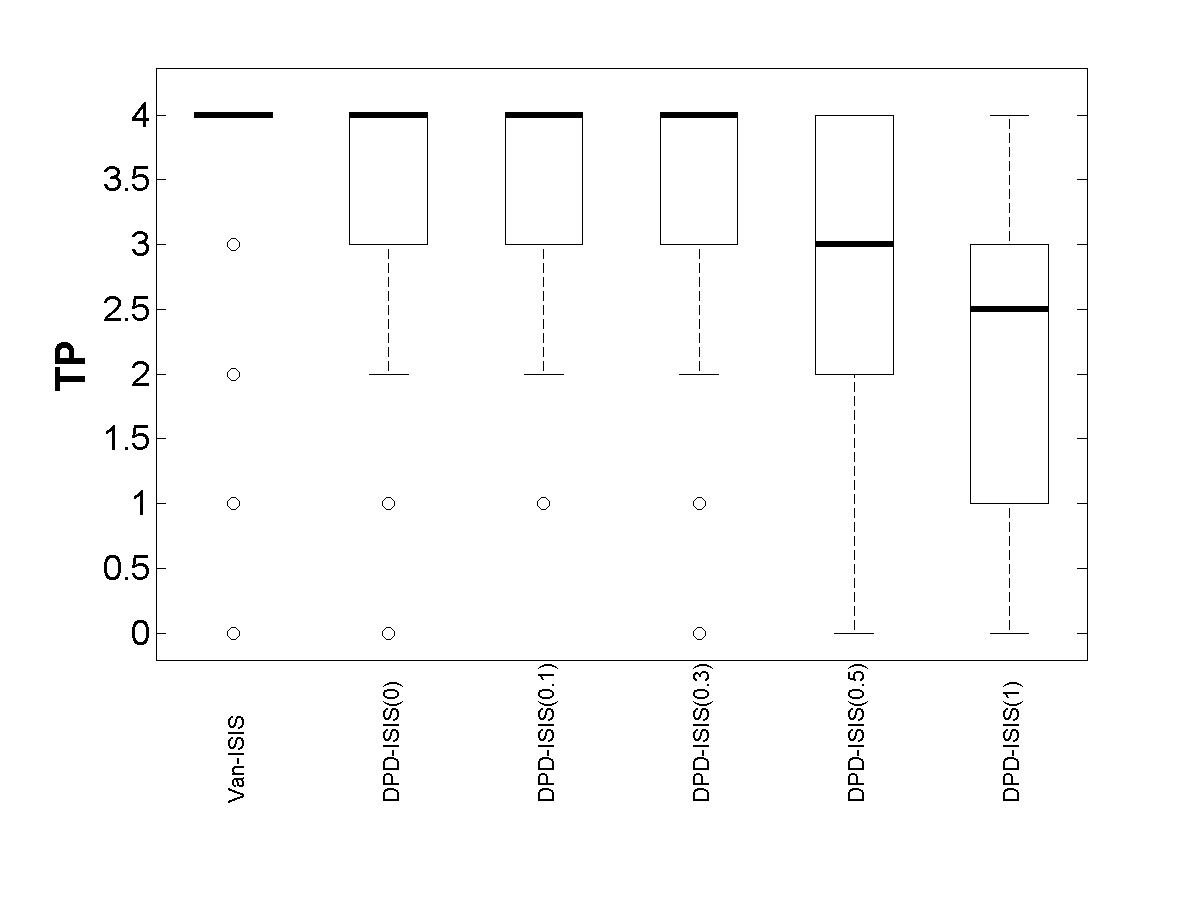}
		\label{FIG:TP_ISIS1_Set3_Ss1_n100_cont00}}
	\subfloat[SNR=1; 5\% contamination]{
		\includegraphics[width=0.33\textwidth]{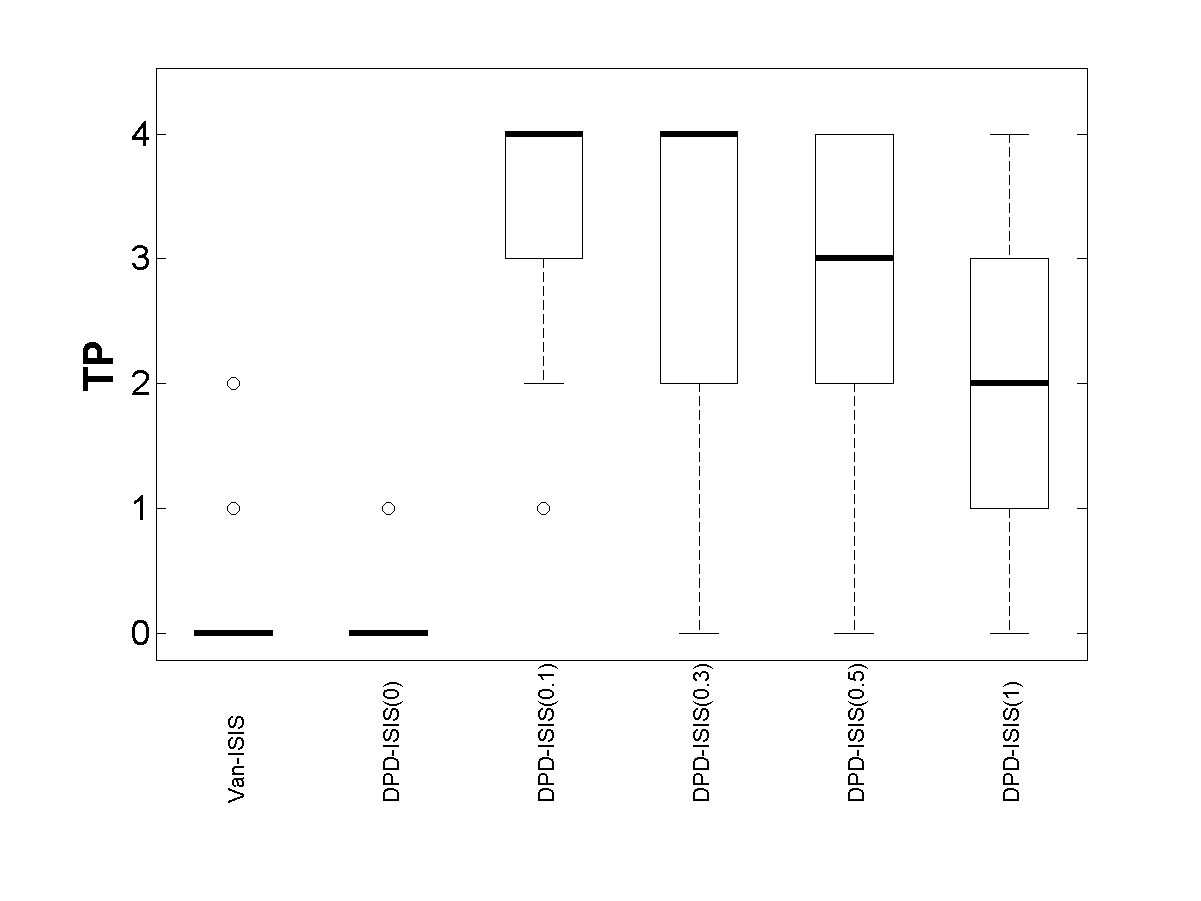}
		\label{FIG:TP_ISIS1_Set3_Ss1_n100_cont05}}
	\subfloat[SNR=1; 10\% contamination]{
		\includegraphics[width=0.33\textwidth]{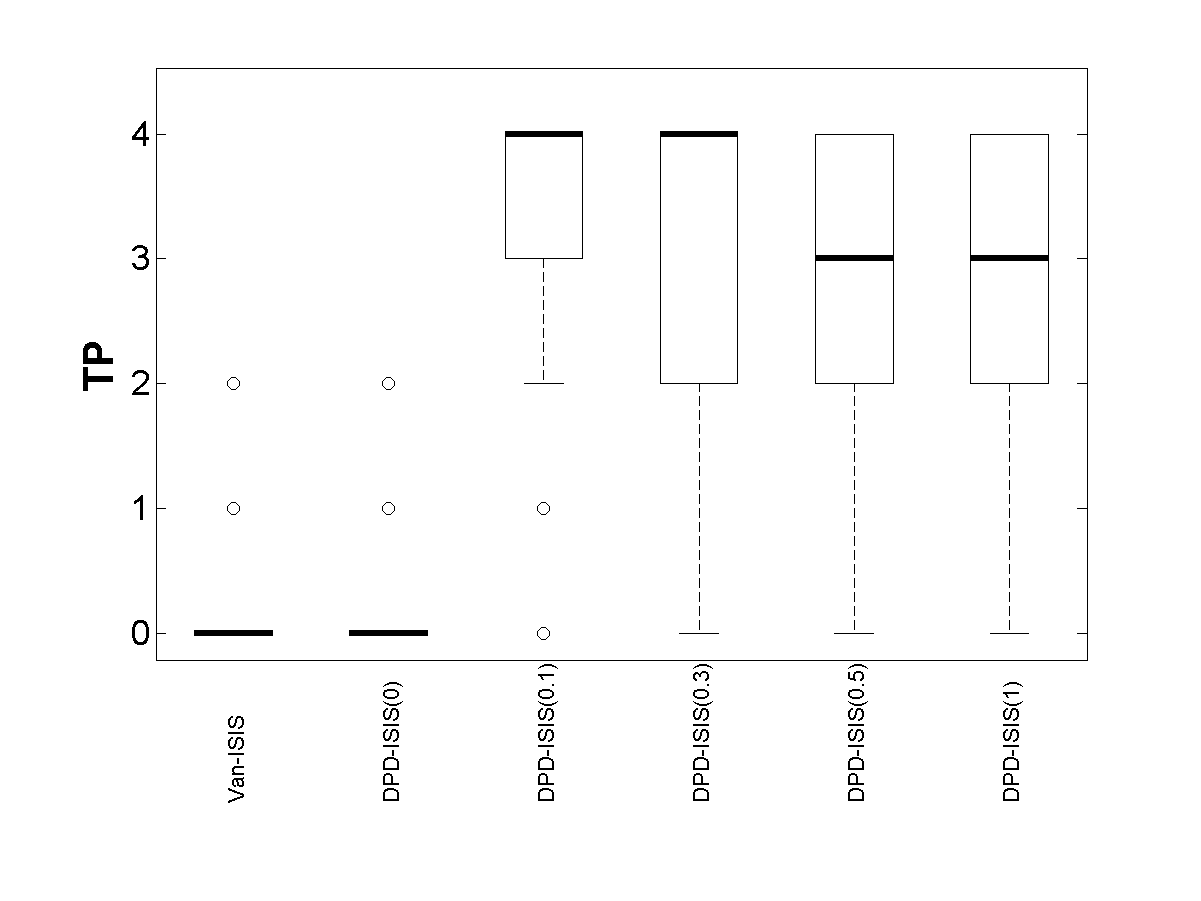}
		\label{FIG:TP_ISIS1_Set3_Ss1_n100_cont10}}
	\\	
\subfloat[SNR=5; No contamination]{
\includegraphics[width=0.33\textwidth]{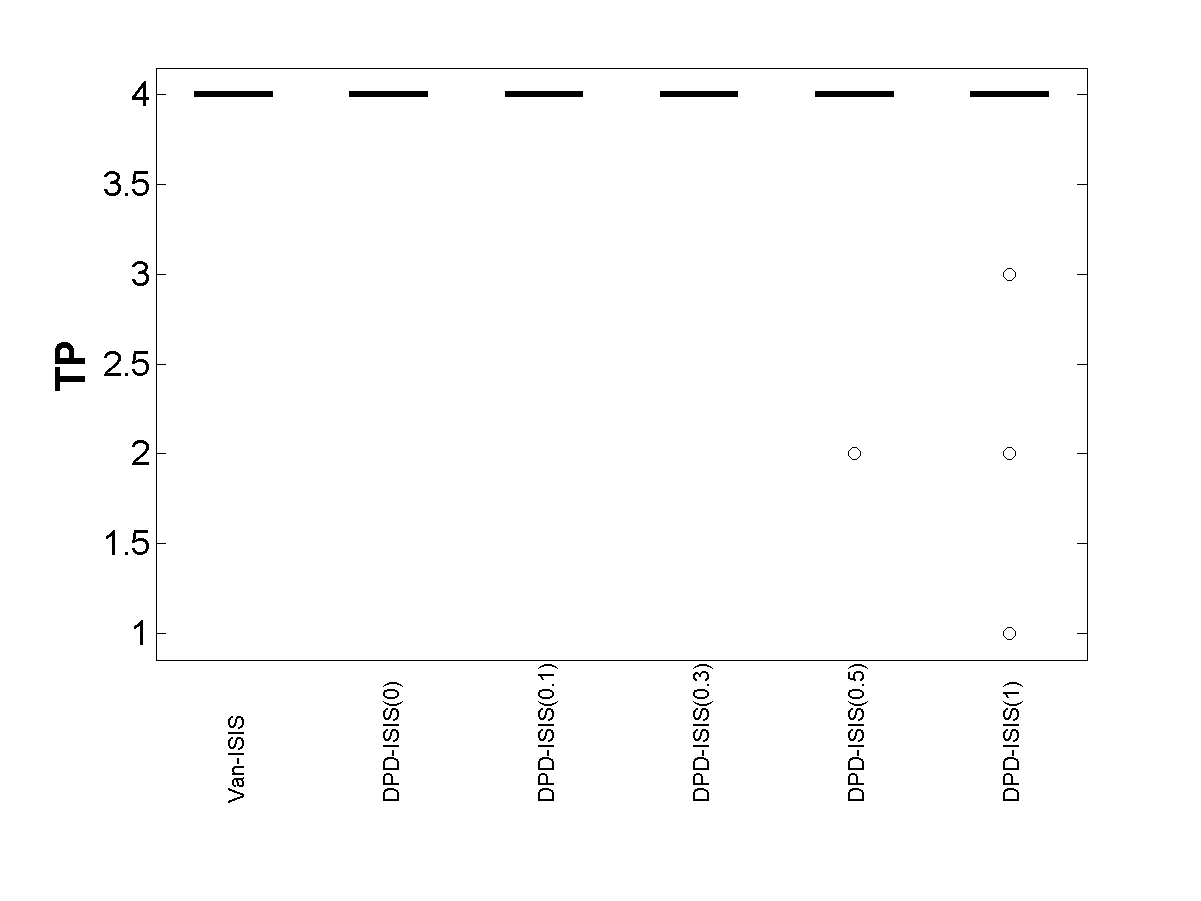}
\label{FIG:TP_ISIS5_Set3_Ss1_n100_cont00}}
\subfloat[SNR=5; 5\% contamination]{
	\includegraphics[width=0.33\textwidth]{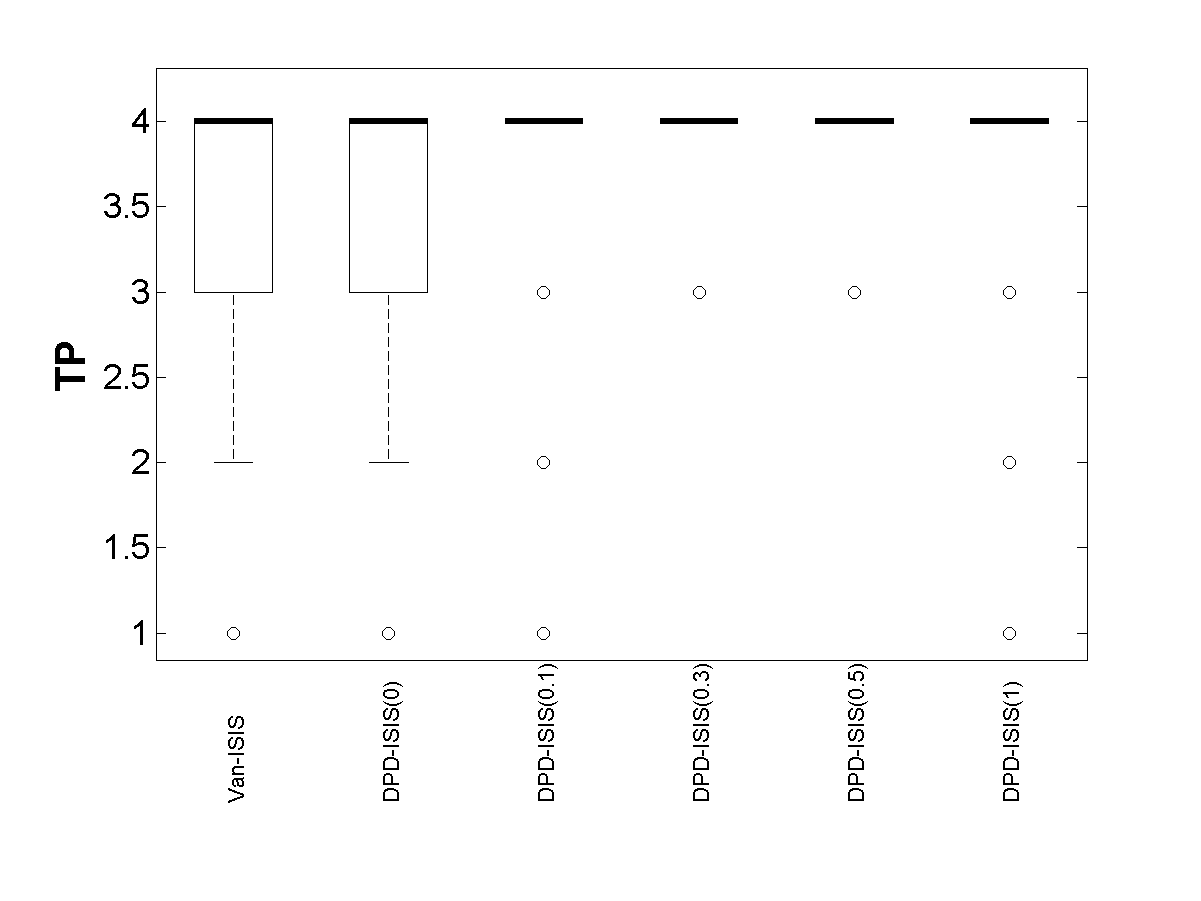}
	\label{FIG:TP_ISIS5_Set3_Ss1_n100_cont05}}
\subfloat[SNR=5; 10\% contamination]{
	\includegraphics[width=0.33\textwidth]{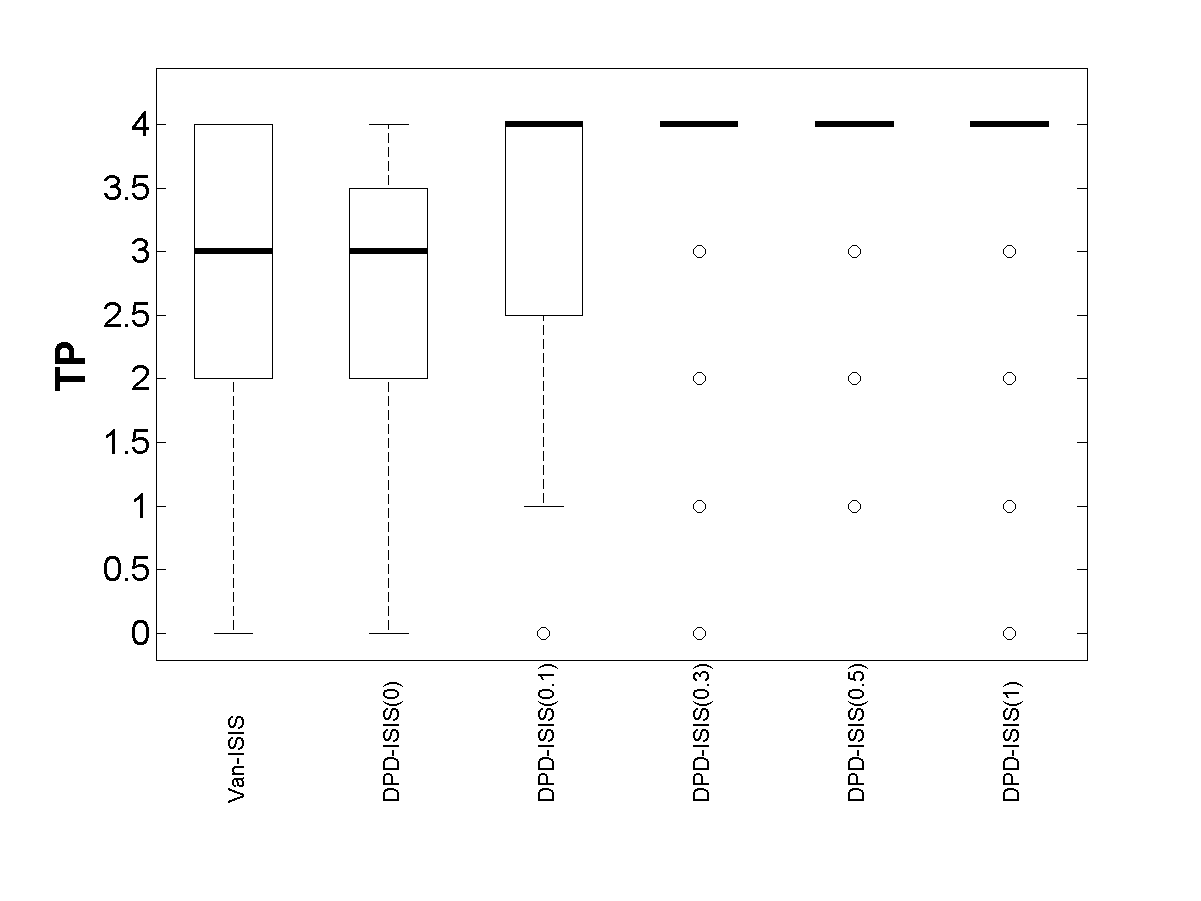}
	\label{FIG:TP_ISIS5_Set3_Ss1_n100_cont10}}
\caption{Box-Plots of the true-positives (TP) obtained by different ISIS approaches with $n=100$ and $p=5000$
for different values of $\tau$ (SNR) and different amount of contamination}
\label{FIG:TP_n100_ISIS}
\end{figure}

We present the average IC values obtained by different DPD-ISIS and Van-ISIS over 100 replications of the above-mentioned simulation
under pure data as well as under contaminated data in Table \ref{TAB:PIC_ISIS}; recall that
it gives us the number of times the correct full model (all the 4 true positives) is selected by the DPD-ISIS approach.
In each replication, as suggested in Section \ref{SEC:DPD_ISIS}, we use the stopping criterion based on the selected active set-size
and the screening is stopped when this active set in a given iteration does not change its size from the previous iteration.
Since the full model is not always selected correctly, 
we also report the individual numbers of true positives selected (TP)  via box-plots in Figure \ref{FIG:TP_n100_ISIS}.

\begin{table}[h]
\caption{Percentage of times the full (correct) model is selected (average IC) by different ISIS approaches for $n=100$ and $p=5000$}%
\centering
\begin{tabular}[c]{|l|r|rrrrr|}\hline
	&	\multicolumn{1}{c|}{Usual }	&	\multicolumn{5}{c|}{Proposed DPD-ISIS($\alpha$)}		\\
	$100\epsilon\%$ &	Van-ISIS	&	$\alpha=0$	&	$\alpha=0.1$	&	$\alpha=0.3$ &	$\alpha=0.5$	&	$\alpha=1$	\\\hline
	\hline
	\multicolumn{7}{|l|}{\underline{Moderate Signal ($\tau=1$; SNR=1)}}\\
0\%	&	79	&	68	&	65	&	58	&	48	&	22 \\
5\%	&	0	&	0	&	63	&	55	&	45	&	20 \\
10\%&	0	&	0	&	51	&	55	&	48	&	31 \\
	\hline
	\multicolumn{7}{|l|}{\underline{Moderate Signal ($\tau=5$; SNR=5)}}\\
0\%	&	100	&	100	&	100	&	100	&	99	&	93 \\
5\%	&	61	&	63	&	93	&	98	&	99	&	89 \\
10\%&	28	&	25	&	68	&	92	&	94	&	92 \\
	\hline
\end{tabular}
\label{TAB:PIC_ISIS}%
\end{table}

From these results along with numerous other simulation studies not presented here, 
it is evident that for any fixed $\alpha$, the DPD-ISIS performs better in identifying more true positives 
as the SNR and $n/p$ ratio increase or the covariate correlation  $\rho$ decreases;
after some limiting values of SNR or $n/p$, 
the DPD-ISIS always can correctly select all the true positives for any amount of correlation among covariates.
This provides an indication that the DPD-ISIS also have similar sure screening property as the usual ISIS under suitably modified conditions.
For any particular case with  no contamination (pure data), in consistence with the DPD-SIS, the performance of DPD-ISIS also deteriorates 
with increasing values of $\alpha$  but such a loss is not significant for smaller values of $\alpha>0$;
under appropriate set-up they all select the correct full model in all the replications yielding 100\% correct results. 

However, under contamination, the performance of the usual Van-ISIS (as well as DPD-ISIS at $\alpha=0$)
becomes significantly worse; under some weaker set-ups, they even fail to select the correct model at all in any replication!
The proposed DPD-ISIS with larger values of $\alpha>0$ yield extremely stable results with very little loss in their performance  
compared to the pure data case. It is also evident that $\alpha\approx 0.3$ provides the best trade-off in the cases considered here;
it always yield the median TP values as the correct active set size (4) with some fluctuations depending on the signal strength 
or amount of contamination in the data.

\subsection{On the Choice of robustness parameter $\alpha$ in DPD-SIS or DPD-ISIS}
\label{SEC:choice}
	
Our proposed DPD-SIS (and also the DPD-ISIS) depends on a tuning parameter $\alpha$ 
which is seen to control the trade-off between the asymptotic efficiency of the underlying MDPDE under pure data 
	and its robustness under contamination.  In terms of variable screening as well, similar trade-offs are observed 
	through our extensive empirical experiments. When there is no contamination in the data, the usual SIS or Reg-SIS 
	(which is DPD-SIS at $\alpha=0$) has the best performance, which deteriorates for DPD-SIS($\alpha$) as $\alpha$ increases
	although the loss is seen to be acceptable for smaller values of $\alpha\leq 0.3$.
	On the other hand, under contamination, the performance of the DPD-SIS becomes more and more stable with increasing values of $\alpha$ 
	while the performance of the usual SIS or Reg-SIS breaks down completely even in presence of small amounts of contamination.
	Considering these trade-offs, it has been observed from our extensive simulation studies that DPD-SIS with $\alpha=0.3$ 
	performs the best under data contamination in all the scenarios considered and it also clearly outperforms all the existing non-parametric methods.
	Based on these experiments, we recommend $\alpha\approx 0.3$ to be a good empirical suggestion to use in most practical applications of DPD-SIS 
	(or DPD-ISIS).

	It is worthwhile to note that, in usual practice with statistical procedure depending on a tuning parameter,
	an adaptive data-driven choice of the underlying tuning parameter is expected and seems to provide the best results in each cases. 
	For the underlying MDPDE used in our DPD-SIS, such data-driven selection procedures for the robustness tuning parameter 
	are available. In the context of linear regression, 
	one such algorithm for selecting the optimal $\alpha$ is explored by \citet{Ghosh/Basu:2015a}.
	However, in the present case of DPD-SIS, we are using MDPDE for each marginal regression model
	and a data-based algorithm will often produce different values of $\alpha$ for each such marginal model,
	since the amount of contamination is often different across covariates. Working with different $\alpha$ values in
	one application of DPD-SIS is not useful and would break the coherence of the analysis -- 
	one should use the same $\alpha$ across all the steps of DPD-SIS in one application to get consistent inference. Additionally, data-driven selection of $\alpha$ would also increase the computation time, which is not an attractive feature in variable screening situations. We believe our empirical suggestion should work well in most applications.

\section{Analysis of Triglyceride Data}\label{SEC:Data_examples}
	
	In this section, we will apply our suggested variable screening method to our motivating example described in the Introduction, 
	and show how this helps us in the variable selection process.
	As described in the Introduction, we have data from 54 individuals who underwent an intervention with intake of capsules of either fish oil, 
	oxidized fish oil or sunflower oil for a period of seven weeks. 
	The study is presented in \citet{Ottestad/etc:2012}. 
	Fasting triglyceride (TG) levels were measured at baseline and after seven weeks of intervention.
	In addition, we have gene expression measured in Peripheral blood mononuclear cells (PBMC). 
	These are immune system cells and because they are circulating cells, 
	they are exposed to nutrients, metabolites and peripheral tissues and may therefore reflect whole-body health. 
	We are interested in relating TG change (seven weeks minus baseline) 
	to gene expressions at baseline and our main goal is to identify genes that may be associated with TG response. Thus, we are primarily interested in variable selection. 
	
	As we have relatively few subjects, outliers might have a profound effect on the result, 
	and hence, we are interested in performing a robust variable screening. 
	Our analysis strategy is as follows: We will perform three iterations of the proposed robust DPD-ISIS (Algorithm 2)
	with RLARS in each iteration (Step 4), 
	followed by a robust $L_1$-penalized regression, 
	the DPD-LASSO method of \citet{Ghosh/Majumdar:2019} to be consistent (in Step 8), to produce our list of selected genes.
	In each iteration of DPD-ISIS we select the $d = n/\log(n) \approx$  13 top variables, 
	while we use penalization parameter $\lambda= \sqrt{\log(p)/n}$ 
	in the final DPD-LASSO. A penalization parameter of this order has been shown to have certain optimality properties, see e.g. page 296, \citet{Hastie/etc:2015}
	We will do this for $\alpha = 0$ (which is not the usual ISIS as discussed in Section \ref{SEC:DPD_ISIS}), 0.1, 0.3 and 0.5 
	and compare the lists of selected genes. 
We have also performed the usual correlation based Van-ISIS \citep{Saldana/Feng:2018} described in Section \ref{SEC:DPD_ISIS} 
as our benchmark of comparison  for the proposed procedures.
When applying Van-ISIS, we observe that the estimated active set size (number of selected genes) does not change after three iterations, 
and we have used this as our stopping criterion. For the sake of comparison, 
we have also performed exactly three iterations of our proposed DPD-ISIS for each $\alpha$.
In the final penalized regression model, we also include treatment group and body mass index. 
However, this does not change the results significantly for any of the procedures. 
We present the results on the number of genes selected in the final model obtained by each procedure in Table \ref{TAB:data1};
the detailed gene list and  estimated regression coefficients in the final model are only presented, for brevity, 
in case of the usual non-robust ISIS (benchmark) and our recommended choice $\alpha=0.3$ in Table  \ref{TAB:data2}.

	\begin{table}[h]
		\caption{Numbers of Genes selected by different ISIS for the Triglyceride data}%
		\centering
		\begin{tabular}{|l|r|rrrr|}\hline
			&	Usual & \multicolumn{4}{c|}{DPD-ISIS with $\alpha$}	\\
			& van-ISIS &	0 & 0.1 & 0.3 & 0.5\\\hline
			Genes selected by ISIS			&	21	& 18	& 26	& 23	& 30 \\
			Genes selected in the final joint model	& 7	 & 9	& 18	& \textbf{21}	& 20 \\
			\hline
		\end{tabular}
		\label{TAB:data1}%
	\end{table}
	\bigskip
	
	\begin{table}[!h]
		\caption{Detailed list of Genes selected by the usual ISIS and the proposed DPD-ISIS
			and associated estimated regression coefficients ($\widehat{\beta}_j$) in the final model 
			for the Triglyceride data}%
		\centering
		\begin{tabular}{|llr|llr|}\hline
			\multicolumn{3}{|c|}{Usual van-ISIS}	&	\multicolumn{3}{c|}{Proposed DPD-SIS($\alpha=0.3$)}\\
			Genes	&	Prob id	&	$\widehat{\beta}_j$	&	Genes	&	Prob id	&	$\widehat{\beta}_j$	\\
			\hline
			HNRNPK	&	ILMN3260017	&	$-$0.004	&	FOXF2	&	ILMN1674934	&	0.081	\\
			NA	&	ILMN1896699	&	0	&	HNRNPK	&	ILMN3260017	&	0.051	\\
			NA	&	ILMN1910805	&	$-$0.042	&	HYAL1	&	ILMN1739813	&	$-$0.399	\\
			NA	&	ILMN1712784	&	$-$0.063	&	UTY	&	ILMN3233091	&	0.309	\\
			NA	&	ILMN1679106	&	0	&	NA	&	ILMN1772136	&	$-$0.227	\\
			NA	&	ILMN1687707	&	$-$0.025	&	RPS27	&	ILMN1660498	&	$-$0.088	\\
			MORC4	&	ILMN1795463	&	0	&	SCARA3	&	ILMN1723358	&	$-$0.058	\\
			RPS16	&	ILMN1651850	&	0	&	SLITRK5	&	ILMN1789040	&	0.060	\\
			ZP3	&	ILMN1672378	&	0.006	&	ZP3	&	ILMN1672378	&	$-$0.427	\\
			FOXF2	&	ILMN1674934	&	$-$0.006	&	ZSCAN12	&	ILMN1786281	&	0.738	\\
			PKLR	&	ILMN1725172	&	0	&	SEZ6L2	&	ILMN2413780	&	$-$0.224	\\
			NA	&	ILMN1881212	&	0	&	FAM161A	&	ILMN3238106	&	$-$0.332	\\
			NA	&	ILMN3242572	&	0	&	EEF1A1	&	ILMN3201843	&	$-$0.091	\\
			TTC8	&	ILMN2401927	&	0	&	ABCD1	&	ILMN3237161	&	$-$0.264	\\
			XCR1	&	ILMN1764034	&	0	&	ALG1	&	ILMN1787954	&	$-$0.197	\\
			TBX1	&	ILMN2248112	&	0	&	AASS	&	ILMN1678323	&	$-$0.248	\\
			PPT2	&	ILMN1750664	&	0	&	EML1	&	ILMN1729455	&	$-$0.152	\\
			KLHL26	&	ILMN1805330	&	0	&	NA	&	ILMN1839740	&	$-$0.313	\\
			SCARA3	&	ILMN1723358	&	0	&	CDCA2	&	ILMN1660654	&	0.163	\\
			NA	&	ILMN1880704	&	$-$0.037	&	NA	&	ILMN1698246	&	$-$0.459	\\
			FGB	&	ILMN2114972	&	0	&	EPB41L4A	&	ILMN1791867	&	$-$0.096	\\
			&		&		&	SLC7A11	&	ILMN1655229	&	0	\\
			&		&		&	TMEM47	&	ILMN2129234	&	0	\\
			\hline
		\end{tabular}
		\label{TAB:data2}%
	\end{table}

	Two observations are worth discussing. 
	First, three times as many genes are selected with the robust procedure (21 vs. 7) as with the non-robust ISIS. 
	Second, there is very little overlap between the two gene sets 
	(only three of the genes selected with $\alpha=0.3$ are selected by van-ISIS).
	If we have a look at the number of genes selected as a function of $\alpha$ (not shown), we observe that the numbers are increasing with increasing $\alpha$, more or less. This is somewhat counterintuitive, as the efficiency of the procedure is reduced with increasing $\alpha$. However, as pointed out earlier, the stability (in terms of robustness) is increasing. We see this as a strong indication of problems with outliers in this rather small dataset, as illustrated in the Introduction. The fact that there is very little overlap between the two gene sets in Table \ref{TAB:data2} can also be seen as an illustration of this problem; with small sample size, outliers are dominating the analysis to a rather large extent.
	It is worth pointing out that the single gene with the strongest effect by DPD-ISIS 
	(ZSCAN12, illustrated in Fig. \ref{FIG:scPlot4}) is not even on the list of selected genes for van-ISIS.

	If we think in direction of biological interpretation of the findings, we observe that a clear majority of the genes have an estimated coefficient with a negative sign, indicating a down-regulation of TG. It should also be commented that three of the identified probes do not map to a known gene, and hence, their function is unclear.

	
	\section{Conclusions}\label{SEC:conclusion}

	In this paper, we have proposed a new robust variable screening procedure for ultra-high dimensional data 
	using the marginal linear regression approach and the minimum density power divergence estimator for the regression parameter. 
	This is extremely important  in modern statistical analyses of large scale data from medical, biological and other applied sciences.
	We have also proposed an iterative version of our DPD based sure independence screening procedure in line with ISIS
	that is helpful for robust variable screening in the presence of correlated covariates. 
	In this paper we have concentrated on linear relationships between the response and all the available covariates
	and hence, our proposed procedure is robust against data contamination (e.g., outliers, or leverage points)
	whenever the assumed linear regression model is approximately correct. The robustness of the proposed DPD-SIS 
	is justified theoretically through use of influence functions and sensitivity analyses and also empirically through an extensive simulation study. 
	It has been empirically shown that the proposed DPD-SIS at suitably chosen robustness tuning parameter $\alpha$ 
	provides the best performance under data contamination and is superior compared to the usual SIS
	as well as several existing robust non-parametric screening procedures under most critical scenarios. 
	We have applied our proposal for the robust analyses of data on triglyceride response to identify the important genes that may cause the variation in triglyceride response between different subjects.

	We have implemented the proposed DPD-based procedure in R for all the simulations and real data analyses.
	The relevant codes are available from the authors upon request, 
	and can be used by any practitioner for robust analyses of their experimental datasets from real-life studies. 
	
	For the practical example it is worth noting that, in order to ensure stability of the final solution, it would make sense to apply some relevant additional procedure, e.g. stability selection \citep{Meinshausen/Buhlmann:2010,Shah/Samworth:2013} on top of the final penalized regression.
	With the promising and encouraging performance of the proposed DPD based robust variable screening procedure, 
	this paper opens up several important directions of future research.
	The first and foremost work would be to prove the theoretical properties of our proposed methodologies, 
	including its claimed sure screening properties.  Also, we have restricted ourselves to linear regression only in this work
	and hence, it would be practically important to extend this robust variable screening procedure 
	to more general parametric regression settings, like generalized linear models.     
	We hope to pursue these important research extensions in our future works.

	
	

\section*{Acknowledgment}
The authors wish to thank Prof. Stine Ulven, Department of Nutrition, University of Oslo, for providing the Triglyceride dataset 
and guiding us in biological interpretation of the results.
A major part of this research work has been done while the first author (AG) was visiting University of Oslo, Norway 
The research of AG is also partially supported by the INSPIRE Faculty Research Grant from Department of Science and Technology, Govt. of India.

\end{document}